\documentclass[sensors,article,submit,pdftex,moreauthors]{Definitions/mdpi} 

\firstpage{1} 
\pubvolume{1}
\issuenum{1}
\articlenumber{0}
\pubyear{2024}
\copyrightyear{2024}
\datereceived{ } 
\daterevised{ } 
\dateaccepted{ } 
\datepublished{ } 
\hreflink{https://doi.org/} 

\usepackage{subcaption}
\usepackage[separate-uncertainty=true]{siunitx}

\DeclareSIUnit\ppm{ppm}
\DeclareSIUnit\Vpp{Vpp}
\DeclareSIUnit\sqrthz{\ensuremath{\sqrt{\text{Hz}}}}
\DeclareSIUnit[per-mode = symbol]\Hzasd{\Hz\per\sqrthz}
\DeclareSIUnit[per-mode = symbol]\masd{\m\per\sqrthz}
\DeclareSIUnit[per-mode = symbol]\fmasd{\femto\m\per\sqrthz}

\Title{A High-Finesse Suspended Interferometric Sensor for Macroscopic Quantum Mechanics with Femtometre Sensitivity}

\TitleCitation{A High-Finesse Suspended Interferometric Sensor for Macroscopic Quantum Mechanics with Femtometre Sensitivity}

\Author{Jiri Smetana $^{1}$*\orcidA{}, Tianliang Yan $^{1}$, Vincent Boyer $^{2}$ and Denis Martynov $^{1}$}

\AuthorNames{Jiri Smetana, Tianliang Yan, Vincent Boyer and Denis Martynov}

\AuthorCitation{Smetana, J.; Yan, T.; Boyer, V.; Martynov, D.}

\address{%
$^{1}$ \quad Institute for Gravitational Wave Astronomy, School of Physics and Astronomy, University of Birmingham, Birmingham B15 2TT, United Kingdom\\
$^{2}$ \quad School of Physics and Astronomy, University of Birmingham, Birmingham B15 2TT, United Kingdom}

\corres{Correspondence: gsmetana@star.sr.bham.ac.uk}

\abstract{We present an interferometric sensor for investigating macroscopic quantum mechanics on a table-top scale. The sensor consists of a pair of suspended optical cavities with finesse over 350,000 comprising \SI{10}{\g} fused-silica mirrors. The interferometer is suspended by a four-stage, light, in-vacuum suspension with three common stages, which allows us to suppress common-mode motion at low-frequency. The seismic noise is further suppressed by an active isolation scheme, which reduces the input motion to the suspension point by up to an order-of-magnitude starting from \SI{0.7}{\Hz}. In the current room-temperature operation, we achieve a peak sensitivity of \SI{0.5}{\fmasd} in the acoustic frequency band, limited by a combination of readout noise and suspension thermal noise. Additional improvements of the readout electronics and suspension parameters, will enable us to reach the quantum radiation pressure noise. Such a sensor can eventually be utilised for demonstrating macroscopic entanglement, and for testing semi-classical and quantum gravity models.}


\begin{document}
/nolinenumbers

\section{Introduction}
\label{sec:intro}

Interferometric devices make excellent candidates for the probing of weak signals on the quantum scale. This is due to the impressive sensitivity achievable in such devices, and their ability to form complex quantum systems.

Interferometers see widespread use in the development of laser technologies, atom trapping~\cite{AtomTraps}, and extend beyond mere photons with atom interferometry used in novel tests of fundamental physics~\cite{AtomInterf,AION}. Laser interferometers make particularly impressive sensors of displacement due to the sharp intensity response induced by microscopic (sub-wavelength) displacements of the key optical components. When one of the mirrors in the interference path is attached to a moving body, its relative displacement can be measured with excellent precision.

One of the most prominent uses of laser interferometry is in the kilometre-scale facilities for detecting gravitational waves (GW). The Advanced LIGO (aLIGO)~\cite{AdvLIGO} and Advanced Virgo (AdV)~\cite{AdvVirgo} gravitational-wave observatories currently serve as the gold standard for precision displacement sensing with aLIGO achieving a peak sensitivity of \SI{2e-20}{\masd} during the most recent completed observing run (O3)~\cite{LIGOSensO3}. Since the original detection of signal GW150914~\cite{BBHDetection}, the detectors have observed dozens of sources~\cite{Catalogue1,Catalogue2,Catalogue3}. Such measurements can give us insight into aspects of general relativity~\cite{GRTest}, stellar and black-hole population statistics~\cite{PopStats} and the structure of neutron stars~\cite{NSProperties}, to name a few. The role of interferometers in GW detection will only continue to expand following construction of the next-generation terrestrial observatories (CE~\cite{CE} and ET~\cite{ET2020}), and the million-kilometre-scale LISA detector~\cite{LISA}. Interferometric detectors are seeing growing application in the search for dark matter particles~\cite{ALPS,DANCE,LIDA}, quantisation of spacetime~\cite{Holometer1,Holometer2}, and entanglement~\cite{Entanglement}, playing a pivotal role in the development of particle physics and cosmology into the future.

For the displacement-sensing laser interferometer, the quantum nature of light imposes an undesirable limitation to the detector's sensitivity. Through the formalism of Caves~\&~Schumaker~\cite{TwoPhoton,Schumaker1985New}, we can analyse the nature of quantum noise in reference to the two so-called quadratures of the electromagnetic field (amplitude and phase), which form a conjugate pair. The intrinsic and inescapable fluctuations in the two quadratures of light couple to the displacement measurement of the interferometer. In a simple Fabry-Perot interferometer, the phase quadrature fluctuations couple directly to the readout, resulting in the so-called quantum shot noise (QSN), whilst fluctuations in the amplitude quadrature induce a force on the cavity mirrors---the so-called quantum radiation pressure noise (QRPN)~\cite{RadPresSource}.

We write the total quantum noise in a suspended, simple Fabry-Perot cavity (zero end-mirror transmission) following the conventions in Ref.~\cite{QuantumBookMiao} as
\begin{equation}
    S_x^{qn}(\Omega) = \left(\frac{1}{\kappa} + \kappa\right) \frac{S_\mathrm{SQL}(\Omega)}{2},
    \label{eq:quantum}
\end{equation}
where $\kappa$ is a frequency-dependent collection of terms given by
\begin{equation}
    \kappa = \frac{\omega_0 T_1 P_c}{m L^2 \Omega^2\left(\Omega^2 + \left(\frac{T_1 c}{4 L}\right)^2\right)},
    \label{eq:kappa}
\end{equation}
and the common parameters are collected in $S_\mathrm{SQL}$, which is given by
\begin{equation}
    S_\mathrm{SQL}(\Omega) = \frac{8 \hbar}{m \Omega ^2}.
    \label{eq:sql}
\end{equation}
In the equations above $\omega_0$ is the laser's central angular frequency, $T_1$ is the power transmissivity of the input coupler, $P_c$ is the power circulating inside the cavity, $m$ is the mass of the cavity mirror, $L$ is the cavity length, and all other symbols retain their usual meaning. The derivation uses the free-mass approximation. Therefore, the equations are valid for suspended interferometers for frequencies $\Omega \gg \omega_n$, where $\omega_n$ is the highest resonance of the suspension. It is readily apparent from Eq.~\ref{eq:quantum} that the quantum noise is minimised when $\kappa = 1$, which gives rise to the concept of the standard quantum limit (SQL) in Eq.~\ref{eq:sql} as a bounding surface corresponding to the minimum noise level achievable in a given interferometer configuration. The SQL, as a trade-off between the light acting as an imperfect probe (QSN) and the light disturbing the probed system leading to back-action (QRPN), was first formulated in Ref.~\cite{BraginskyQuantum}. An important observation is the simplicity of Eq.~\ref{eq:sql}, with the SQL boundary being only dependant on the mass of the cavity mirrors.

However, the SQL does not represent a fundamental quantum limit to any and all interferometric systems. The development of nonlinear optics has enabled us to manipulate the quantum noise components in the two quadratures in the form of optical squeezing---a technique that reduces the quantum noise in one quadrature at the expense of a rise in the other, first proposed in Ref.~\cite{SqueezingIntroduction}. As the interferometer's displacement signal is read out from only one quadrature, shifting the quantum noise from this quadrature to the other has the effect of reducing the measured quantum noise to a level below the nominal SQL defined above. Squeezing has been demonstrated across a range of experiments~\cite{SqueezingExp1,SqueezingExp2,SqueezingFreqDep} and successfully applied in the current generation of GW detectors~\cite{SqueezingLIGO,SqueezingVirgo}. Other techniques have been proposed for manipulating and reducing quantum noise, such as phase-insensitive amplification via opto-mechanical coupling~\cite{QuantumAmpOriginal,QuantumAmplifierTD,QuantumAmpIdeal}.

Whilst the theory of the quantum behaviour of light is well advanced, many aspects of the theory lack experimental verification. Crucially, no experiment has yet operated at the sensitivity level of the SQL, with the GW detectors unable to reach the SQL due to the coating thermal noise~\cite{LIGOSensO3,VirgoSens}. However, an interferometer operating at the quantum noise level over a broad band covering the SQL frequency would enable us to investigate novel aspects of quantum behaviour, such as entanglement on the macroscopic scale~\cite{MacroscopicEntanglement} and allow us to test the quantum nature of gravity~\cite{SemiGrav,QuantumGrav}.

In this paper, we present our table-top interferometeric sensor for probing macroscopic quantum mechanics and demonstrate a femtometre peak displacement sensitivity in the audio band. The experiment closely follows practices and experience adopted from the GW community motivated by the advanced noise suppression techniques implemented in GW detection and the near-SQL sensitivity of the detectors. The interferometer consists of a high-finesse suspended cavity in a cryogenic environment. Suspending the mirrors (a feature we share with GW detectors) is a departure from typical tabletop configurations but is beneficial for our purposes as the low resonant frequencies of the suspension chain increase the coupling strength of the QRPN to displacement, and thus allow us to eventually reach the SQL.

We design the system to place the SQL frequency at 100 Hz, consistent with the aLIGO detectors, but raise the SQL level by reducing the mass of the cavity mirrors to \SI{10}{\g}. Further reduction of mass is not necessarily desirable due to the likely increase in the suspension resonances (cantilever microresonators, for example, can have a resonant frequency in the kHz to 100s kHz~\cite{Microresonators}) and the increased control difficulty arising from the Sigg-Sidles instability~\cite{SiggSidles}. This mirror mass also has the benefit of corresponding to the typical mass of common 1-inch tabletop optics. A similar approach of using gram-scale mirrors in a suspended cavity to reach the QRPN has been investigated in Ref.~\cite{Wipf} and a large-scale experiment for reaching the SQL is currently undergoing commissioning at the Albert Einstein Institute (AEI) 10-metre prototype~\cite{AEI10m}. This work follows on from preliminary presentations in Refs~\cite{ConferenceSQL1,ConferenceSQL2} and the design study in Ref.~\cite{SmetanaThesis}.

We explore in detail the key aspects of the design in Sec.~\ref{sec:experiment}, focusing on the optical system, the internal suspension and the active isolation system. In Sec.~\ref{sec:results}, we present the sensitivity of the sensor together with a discussion of the noise budget and future steps towards reaching QRPN-dominated operation---the next major milestone towards realising a full quantum sensor.

\section{Experimental Layout}
\label{sec:experiment}

The two key noise sources that limit such systems from reaching the SQL are the thermal noise and seismic noise. The latter is mitigated by the choice to suspend the cavity with a sufficiently soft multi-stage suspension. To this end, we have the examples from across the GW community to draw upon (see Section~\ref{subsec:sus}). The thermal noise is the most challenging noise source to mitigate, best reflected by the aLIGO and AdV detectors, which, whilst close to the SQL, are currently dominated at the SQL frequency by the coating thermal noise~\cite{LIGOSensO3,VirgoSens}.

Via the fluctuation-dissipation theorem~\cite{FlucDiss1,FlucDiss2,FlucDiss3}, we can write the thermal noise of a generic mechanical system~\cite{Saulson} as
\begin{equation}
    S_x^{th}(\Omega) = \frac{4 k_B T \mathfrak{Re}(Y(\Omega))}{\Omega^2},
    \label{eq:fluc_diss}
\end{equation}
where $Y(\Omega)$ is the mechanical admittance and contains the dissipative element of the system's response. The most direct strategy for reducing thermal noise, as expressed in Eq.~\ref{eq:fluc_diss}, is simply to reduce the temperature of the system, hence we design our experiment around the constraints of a cryostat to enable future cryogenic operation. Aside from temperature, the specifics of the system's response are crucially important and the key component of thermal noise mitigation is the appropriate choice of material based on desirable mechanical and thermal properties. Our approach shares some of the cryogenic challenges with the GW detector, KAGRA~\cite{Kagra}, and prospective next-generation detectors, LIGO Voyager~\cite{Voyager}, and ET, which we draw upon in our design.

One important consequence of cryogenic operation is the increased difficulty of implementing active components. Even at the current room-temperature stage, we design the optical and mechanical system without any in-vacuum suspension sensing or active control. If we can achieve low-noise operation in this regime, we will avoid significant technological challenges down the line.

\begin{figure}[H]
    \begin{adjustwidth}{-\extralength}{0cm}
    \centering
    \begin{subfigure}{0.4\linewidth}
        \centering
        \caption{}
        \vspace{2mm}
        \includegraphics[angle=-90,origin=c,width=\linewidth]{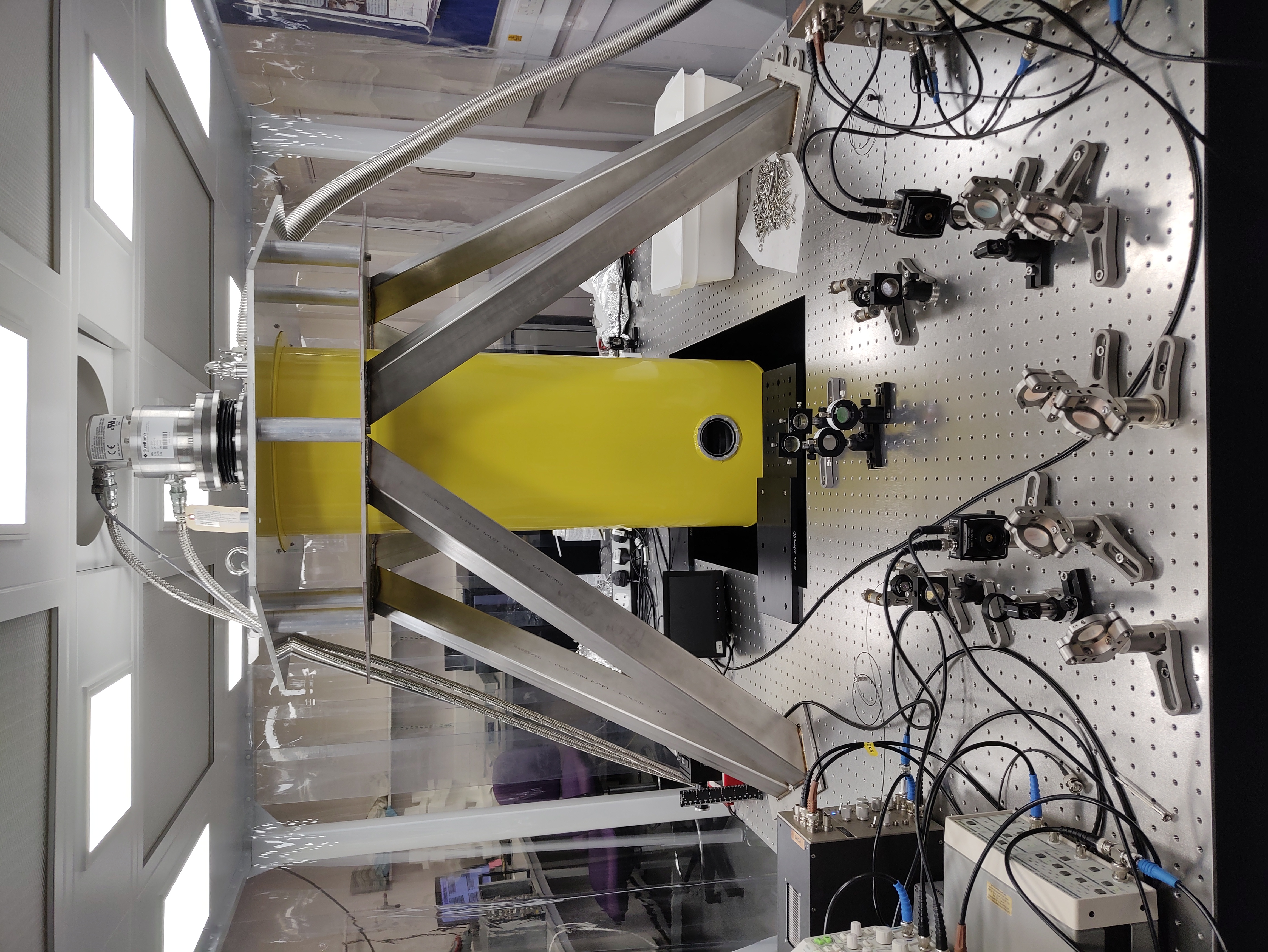}
        \label{subfig:photo-orig}
    \end{subfigure}
    \hfill
    \begin{subfigure}{0.58\linewidth}
        \centering
        \caption{}
        \vspace{2mm}
        \includegraphics[width=\linewidth]{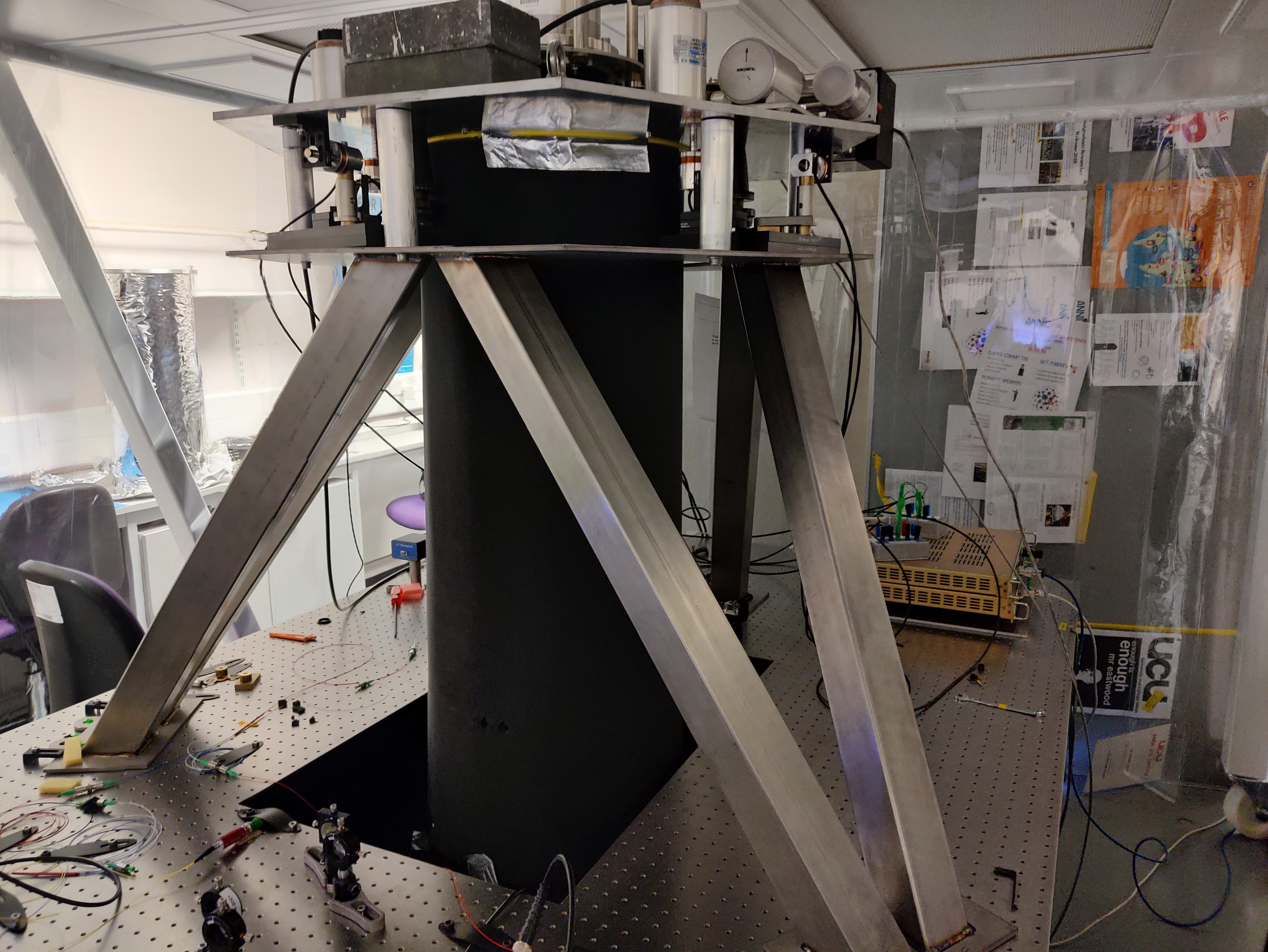}
        \label{subfig:photo-new}
    \end{subfigure}
    \end{adjustwidth}
    \caption{Panel (\textbf{a}) shows the experimetal set-up in an earlier iteration without active seismic isolation of the cryostat. This image shows the key optical and electronic components, as well as the exposed vacuum lid showing the location of the viewports. Panel (\textbf{b}) shows the same experiment in its current status. The parallel plates near the top of the cryostat support an array of seismometers, coil-magnet actuators, passive rubber springs and counterweights. In this image, many electronic components have been moved and the cryostat lid has been wrapped in soft foam, thus obscuring most of the viewport's aperture. These latter changes were done to address acoustic noise that we discovered coupled into some areas of the detection band.}
    \label{fig:photos}
\end{figure}

\subsection{Optical System}
\label{subsec:optical}

\begin{figure}
\centering
\includegraphics[width=0.5\textwidth]{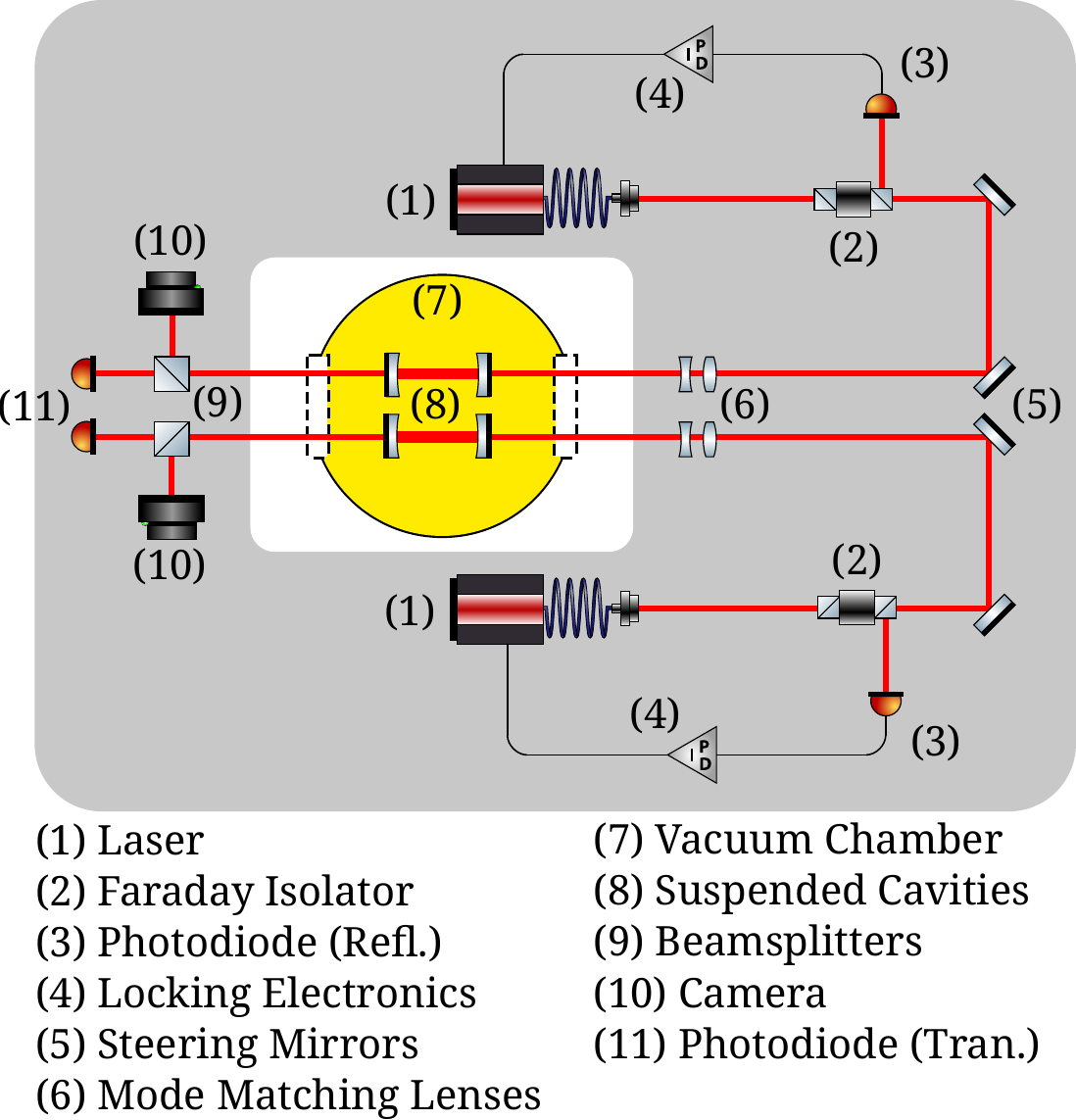}
    \caption{Optical layout of the experiment showing the two identical interferometric systems and the suspended cavities inside the vacuum envelope. Figure reproduced from Ref.~\cite{SmetanaThesis}.}
\label{fig:optical}
\end{figure}

The optical layout is duplicated into two identical systems each based around an independent, suspended Fabry-Perot optical cavity. The cavity mirrors are the only optical components that require cooling and isolating and are, therefore, the only optical components inside the vacuum envelope. The optical layout is shown in Fig.~\ref{fig:optical}.

We currently use fused-silica mirror substrates in the room-temperature iteration of the experiment due to their good optical and mechanical properties (low absorption~\cite{SilicaAbsorption}, loss angle \num[print-unity-mantissa=false]{e-7}~\cite{VariousSilica}). Silica performance degrades at low temperatures due to the loss peak of \num[print-unity-mantissa=false]{e-3}~\cite{SilicaPeak} around \SI{20}{\kelvin}, making its use unsuitable for later cryogenic operation. In the future we plan to transition to silicon substrates due to the superior mechanical properties---loss angle of \num[print-unity-mantissa=false]{e-7}~\cite{Cryogenics}. Its worse thermoelastic noise due to the relatively high coefficient of thermal expansion is mitigated by appropriately tuning the operating temperature near the critical temperature of \SI{20}{\kelvin} where the thermal expansivity crosses zero~\cite{SiliconExpansion}.

The dielectric coatings we use currently are composed of Ta$_2$O$_5$/SiO$_2$ layers deposited by ion beam sputtering~\cite{IBS}. These are chosen due to their high quality (low optical losses~\cite{Absorption} and loss angle in the range \numrange{1e-4}{4e-4}~\cite{CoatingThermal}) and proven track record in precision interferometry. In preparation for the switch to silicon substrates, we also consider other coating options more suitable for cryogenic operation. We currently consider amorphous---a-Si\footnote{amorphous silicon}/SiO$_2$, a-Si/SiN--or crystalline---GaAs/AlGaAs---candidates.

We measure the bandwidth (FWHM) to be \SI{4020}{\Hz} for a cavity length of \SI{95}{\mm}. This yields a measured finesse of \num{3.9e5}. The measured transmissivity of the mirrors is $7.5^{+1.0}_{-1.5}$\,ppm. From the measured finesse, we determine the total round-trip losses to be approximately \SI{16}{\ppm}, which suggests that the additional cavity losses on top of the transmissivity do not exceed more than a few ppm. However, due to the difficulty of measuring the individual loss components (including the mirror transmissivity) to sub-ppm precision, we cannot provide a comprehensive breakdown of each source of loss. Our estimate of the cavity loss is generally in agreement with the study in Ref.~\cite{RealisticCavities}, which reviews the realistic prospects for low-loss optical cavities of different lengths. This results in a high build-up of intra-cavity power, which amplifies the QRPN coupling without the need for high input power. Such high finesse coupled with the softly suspended and light cavity mirrors presents significant challenges to controllability.

The combined challenge of the above factors is most noticeable during the initial lock acquisition. We make use of a Pound-Drever-Hall locking scheme~\cite{PDHScheme}, with sidebands at \SI{4}{\MHz}. Due to the lack of active cavity control, we are limited to locking via laser-frequency actuation. We use a Rio Orion \SI{1550}{\nm}, \SI{24}{\mW} distributed feedback laser with two means of frequency control: the slow but large range control of the laser temperature, and the fast but low-range modulation of the laser current supply. The current controller is further divided into a slow DC controller with DC range of \SI{800}{\MHz} frequency tuning over \SI{8}{\Vpp}, and an AC-coupled fast controller with a peak response at \SI{20}{\kHz}, continuing into 10s MHz, and a range of \SI{400}{\MHz} frequency tuning over \SI{1}{\Vpp}. Due to the short cavity length, the cavity free-spectral range (FSR) is \SI{1.67}{\GHz}, which is outside of the current controller range and makes lock acquisition more challenging.

The resonant condition is maintained via feedback to the two supply current channels with a servo bandwidth of \SI{300}{\kHz}. This is necessary, as the high finesse leads to a narrow error signal, which, combined with the relatively high laser frequency noise of \SI{10}{\Hzasd} at \SI{1}{\MHz}, places strict requirements on the servo bandwidth. The temperature setpoint can be tuned periodically by hand to compensate for accumulated laser frequency drifts on the week time scale.

Another challenge of the lock acquisition process is the initial buildup of intra-cavity power. This generates strong classical radiation pressure, which causes a large drift in the cavity length due to the soft suspension and light cavity mirrors. With power in excess of \SI{0.5}{\m\W}, this shift is beyond the range of the current controller and the cavity drifts out of lock. Our current strategy for achieving lock is through the use of a variable optical attenuator (VOA). The VOA allows us to maintain low input power initially, which reduces the magnitude of the rapid radiation pressure buildup during initial lock. Once a stable lock is established, we can increase the input power slowly whilst compensating for the frequency shift with the laser temperature to keep within the range of the current controller.

We set up two identical systems and beat the two independent lasers together to form an additional readout channel. This is necessary as we require a readout channel that is free of the laser frequency noise, which lies 6 orders of magnitude above the nominal SQL sensitivity at \SI{100}{\Hz}. The issue could be resolved with a single cavity and a low-noise frequency reference but this is a challenge given our strict noise requirements at \SI{100}{\Hz}. The aggressive seismic noise suppression above \SI{10}{\Hz} coupled with the laser-frequency locking to the cavity motion means that the laser phase is very stable in the key frequency band around \SI{100}{\Hz}. The cavity can, in effect, act as a high quality frequency reference in a limited band. Introducing a second cavity with very similar noise characteristics allows us to use one cavity as the frequency reference for the other.

We can tune the central frequency of the beat-note by shifting the lock-point of each cavity by integer multiples of the FSR to a coarse value around \SI{150}{\MHz}. We then measure the frequency drift of this beat-note with a custom-built phase lock loop (PLL) based around a Minicircuits JCOS-175LN voltage-controlled oscillator (VCO). The key challenge here is one of dynamic range. Whilst the two cavities are quiet within the frequency band of interest, they suffer from high root-mean-square (RMS) motion, which comes almost entirely from the seismically excited mechanical resonances of the suspension chain. In order to achieve a sufficiently low electronics noise of the PLL, we use a lower VCO frequency range, which reduces the maximum tolerable cavity RMS motion. The natural RMS motion is larger than this VCO range, which means that we cannot reach our noise requirements with a purely passive seismic isolation system. Our solution involves the active inertial control of the cryostat, which is discussed in more detail in Sec~\ref{subsec:active}.

The locking procedure is entirely achieved using analogue electronics. However, we digitise the control signals for processing using the Control and Data System (CDS) used in aLIGO. The input signals are digitised at \SI{64}{\kHz} by a 16-bit, \SI{20}{\Vpp} analogue-to-digital converter (ADC) via an input anti-aliasing filter with a cut-off frequency at \SI{7}{\kHz}. This allows us to easily search for coherence between different signals during commissioning and is also used to shape the control signals of the low-bandwidth active isolation feedback loops.

\subsection{Suspension}
\label{subsec:sus}

Seismic noise comprises one of the principal sources of noise in suspended interferometers and is generally the dominant noise source at low frequencies. The conventional means of suppressing this noise within the GW community is to softly suspend the optics to achieve a sufficient level of passive isolation. Some different approaches to high-performance suspension design can be seen in the aLIGO quadruple suspension~\cite{LIGOQuad}, the AdV \textit{superattenuator}~\cite{VirgoSuperattenuator}, and the predecessor GEO600 suspension~\cite{LocalControl}.

To achieve horizontal isolation, we use a four-stage pendulum chain formed of intermediate aluminium masses and steel-wire connections. The suspension is attached to the \SI{10}{\K} cold plate inside the cryostat to facilitate conductive cooling of the suspension and cavity optics in the future. Aluminium is chosen for its light weight (see below) and good thermal properties, particularly the impressive thermal conductivity of high purity aluminium (in excess of \SI{e4}{\W\per\m\per\K} at cryogenic temperatures~\cite{Cryogenics}. The steel wires are an intermediate step for room-temperature operation. The poor conductivity and relatively high loss angle of steel makes it unsuitable for cryogenic operation for both cooling and thermal noise considerations. Fused silica and silicon fibres are possible candidates for later cryogenic operation due to their low mechanical loss, whilst beryllium copper or tungsten show promise from a thermal conductivity perspective. The upper three stages are suspended over a length of \SI{19}{\cm} each with the cavity mirrors suspended by a shorter two-wire pendulum of length \SI{2}{\cm}.

For vertical isolation, we use triangular blade springs. The blades have a base width of \SI{3}{\cm} and a length of \SI{130}{\cm} and \SI{200}{\cm} for the top stage and intermediate stages respectively. The thicknesses are tuned to lower all vertical resonances below \SI{10}{\Hz}. The intermediate stage blades are arranged along the diameters of their respective stage masses to maximise the length, as a longer and thicker blade achieves a lower resonant frequency than a shorter and thinner blade given equal internal stress (Section~6.4 in~\cite{BladeEquationTriangle}). There are three vertical stages (one fewer than horizontal) on account of the stronger direct coupling of horizontal motion to the cavity length in contrast to the lower cross-coupling of vertical-to-horizontal motion. We currently use 316 stainless steel due to availability limitations, which has a comparatively low yield strength of around \SI{200}{\MPa} compared to the potential yield strengths in excess of \SI{1}{\GPa} of numerous maraging steels~\cite{MaragingYield}. With a better choice of material in the future, it will be possible to lower all vertical resonances to below \SI{3}{\Hz}.

We use eddy current damping~\cite{EddyDamping} between the upper-intermediate and penultimate masses to avoid bypassing the upper two stages of isolation. The purpose of the damping is to reduce the Q factors of the suspension resonances and thus reduce the RMS motion and ease the range requirements on the PLL outlined in Section~\ref{subsec:optical}. For efficient damping of all resonances from a single stage, it is important to ensure all stages are well coupled to each other. This is achieved by making all stages comparable in mass and length, which is difficult to realise given the fixed 10 g mass of the mirrors. As such, the passive damping is only a partial solution to the RMS problem. Alternative damping solutions exist, such as active damping as is done with the Birmingham optical sensor and electromagnetic motor (BOSEM)~\cite{Bosem} in aLIGO, which offer certain advantages, such as preventing a factor of $f$ loss of suppression at high frequencies. We, however, will continue to avoid active in-vacuum components unless deemed absolutely necessary.

\begin{figure}
\centering
\includegraphics[width=0.8\textwidth]{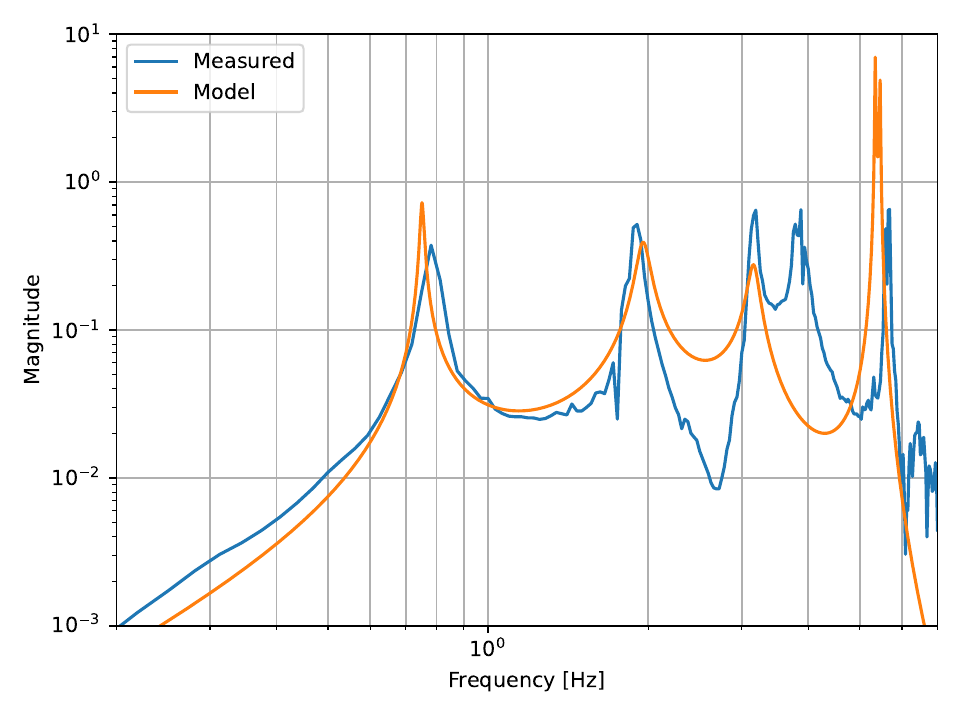}
    \caption{Transfer function of the suspension-point motion to the differential cavity motion. The Lagrangian model uses independently obtained values for the suspension parameters and only contains one free parameter (aside from the Q of the resonances), which is the unknown mismatch between the wire stiffness of the two final mirror stages. The model can be further improved by constraining the clamping losses and other damping mechanisms to better predict the resonance Q factors, and by including the coupling between the two separate cavities (the current model only considers a single cavity), which would likely allow us to explain the zero in the transfer function at \SI{2.7}{\Hz}. The modelled transfer function has been shifted vertically to line up with the measured data for visual clarity. The true gain of the transfer function is determined by parameters that are difficult to measure independently and is thus unknown. However, the location of the resonances and the overall shape of the transfer function is independent of the overall gain and these match the measured data with reasonable accuracy. The unknown resonance at approximately \SI{3.8}{\Hz} is thought to be a pitch resonance that has cross-coupled to the the longitudinal mirror displacement.}
\label{fig:tf-cryo-cav}
\end{figure}

An important aspect of the suspension design is that both cavity mirrors are suspended from the same common penultimate mass. This provides common-mode rejection of seismic noise in the cavity displacement spectrum below the final mirror-stage resonances, as the cavity is only sensitive to differential motion. This leads to an $f^2$ suppression of seismic noise towards DC below the first suspension resonance. The gain of the suspension-point-to-differential-cavity transfer function is proportional to the mismatch between the stiffness of the two final mirror stages, meaning it is important to ensure the two suspensions are as similar as possible. In order to minimise the mass of the intermediate stages (for efficient damping) but maximise the moment of inertia to reduce angular resonances, we arrive at a ring-like structure. The full suspension schematic is shown in Fig.~3.9 of Ref.~\cite{SmetanaThesis}. The suspension model is based on a Lagrangian analysis, which is also presented in detail in Ref.~\cite{SmetanaThesis}. In Fig.~\ref{fig:tf-cryo-cav} we show the transfer functions of displacement at the suspension point to differential cavity displacement for the principal cavity-axis-aligned horizontal degree of freedom (DoF, referred to in this case as the X DoF). This figure contains the measured transfer function, achieved by exciting the X DoF with the active isolation's actuators (see below) and witnessing with the cavity-locking control signal. The measured transfer function is compared with the result of the Lagrangian analysis and correctly reproduces the overall shape of the transfer function, including the gain and the location of the resonances.

\subsection{Active Isolation}
\label{subsec:active}

The beat-note RMS must be reduced to below the VCO range of \SI{50}{\MHz}, which we cannot achieve with passive internal isolation alone. As the RMS is dominated by seismic coupling to the cavity lengths, we equivalently need to reduce the RMS seismic noise to significantly below \SI{20}{\nm}. We achieve this by actively controlling the cryostat motion. This technique of inertial isolation is widely used within the GW community. We adopt principles shared with the control of Stewart platforms~\cite{StewartPlatform}, inertial sensors, such as the 6D six-axis seismometer~\cite{New6D}, and the high-performance isolation provided by the internal seismic isolation (ISI) of aLIGO~\cite{LIGOSeismicStrategy}.

The general scheme consists of an inertial sensor, actuator and a sufficiently mechanically compliant support structure. Sensing is achieved with six single-axis L-4C geophones (three horizontal and three vertical). The signals from the spatially distributed and strategically oriented geophones can be combined to yield a diagonalised measurement of the full six DoFs (three linear and three angular).

The mechanical structure consists of a rigid cryostat frame, the \SI{140}{\kg} cryostat payload (of which the internal suspension and cavity comprises a negligible amount) and a set of three natural-rubber cone feet that the payload rests upon. The truncated rubber cones are \SI{50}{\mm} and \SI{37}{\mm} diameter at the base and top respectively, with a height of \SI{62.5}{\mm}. This follows a similar strategy to the use of Viton support feet demonstrated in the passive isolation of a multi-stage optical table~\cite{Viton}. The relatively low Young's and shear modulus of rubber leads to low resonant frequencies in all degrees of freedom, with horizontal resonances of \SI{3.9}{\Hz} and a vertical resonance at \SI{7}{\Hz}. Such a soft system would not be suitable for our use in a purely passive capacity. Whilst the low resonant frequencies improve suppression of seismic vibrations from the ground, they also increase the mechanical admittance to the force generated by the pulsing cryopump, which would degrade the overall performance during cryogenic operation. The benefit of the active system is that the cryostat is inertially isolated from vibrations coupling through both channels.

We actuate on the cryostat relative to the rigid cryostat support frame with coil-magnet actuators derived from the BOSEM (sensing components removed). The wire coils have a resistance of \SI{41.4}{\ohm} and an inductance of \SI{17.8}{\milli\H}.

\begin{figure}[H]
    \begin{adjustwidth}{-\extralength}{0cm}
    \centering
    \begin{subfigure}{0.48\linewidth}
        \centering
        \caption{}
        \vspace{2mm}
        \includegraphics[width=\linewidth]{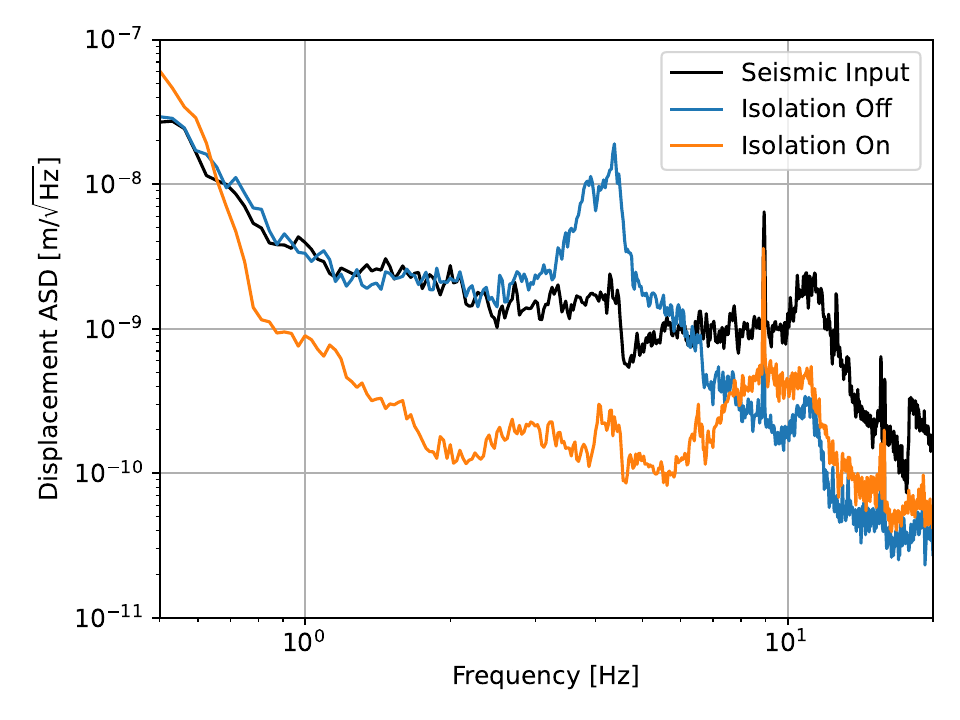}
        \label{subfig:iso-geo}
    \end{subfigure}
    \hfill
    \begin{subfigure}{0.48\linewidth}
        \centering
        \caption{}
        \vspace{2mm}
        \includegraphics[width=\linewidth]{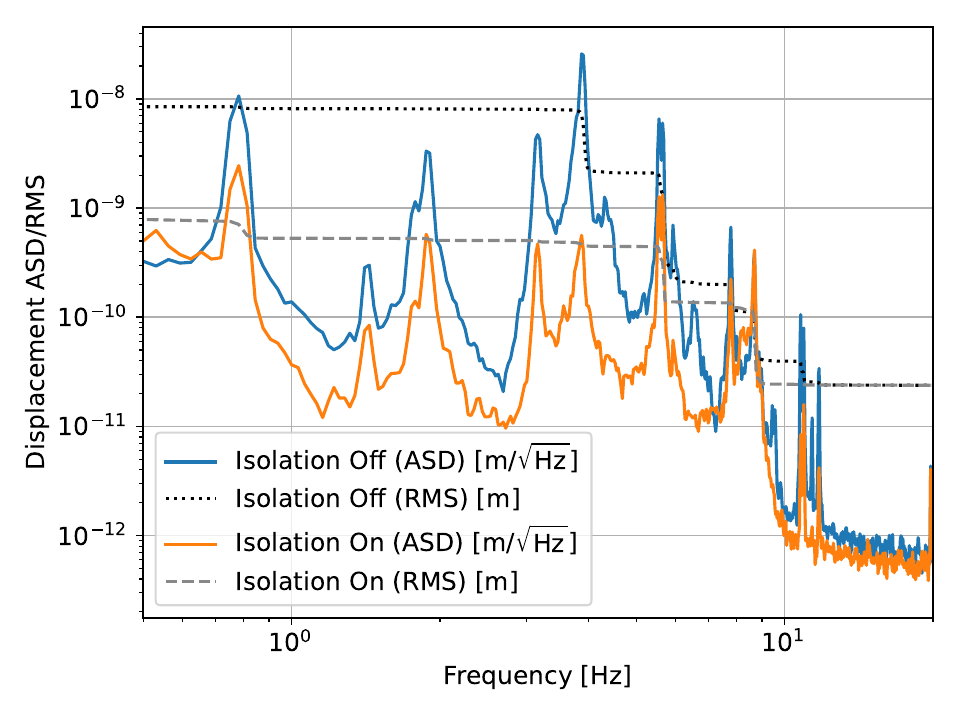}
        \label{subfig:iso-cav}
    \end{subfigure}
    \end{adjustwidth}
    \caption{Panel (\textbf{a}) shows the performance of the active seismic isolation as seen by the geophones in the X DoF. The \textit{isolation off} trace clearly shows the presence of the horizontal resonance of the rubber supports at \SI{3.9}{\Hz} and the passive suppression effect that is generated above this frequency. The \textit{isolation on} trace shows an order-of-magnitude level of suppression compared to the input seismic motion and significant improvement over the passive-only effect in the region below \SI{8}{\Hz}. The \textit{isolation on} trace inherits the passive isolation of the rubber springs above \SI{8}{\Hz}. Panel (\textbf{b}) shows the suppression of the seismic noise by the active isolation as witnessed by the cavity control signal. The RMS curves show the cumulative RMS integrated from high to low frequencies. The comparison of the \textit{isolation on/off} curves shows a consistent reduction in the displacement spectrum with that measured by the in-loop geophones in Fig.~\ref{subfig:iso-geo}. Importantly, the cavity sees an order-of-magnitude drop in its RMS motion when the active isolation in engaged.}
    \label{fig:isolation}
\end{figure}

We make use of the aforementioned CDS for data acquisition and control, which enables us to design a custom digital servo for the feedback control and make coherence measurements with the cavity length readout and obtain an out-of-loop witness channel. This allows us to verify that the inertial isolation of the cryostat is truly leading to a reduction in the seismic coupling to the cavity motion. We construct a feedback servo, which leads to an order-of-magnitude reduction in the RMS displacement of the X DoF (see Fig.~\ref{subfig:iso-geo}). This is reflected in a corresponding order-of-magnitude reduction in the differential cavity motion (Fig.~\ref{subfig:iso-cav}). This suppression leads to a reduction of the RMS cavity displacement to below 1 nm, which is well within the range of the VCO with a sufficient safety margin.

\subsection{Results}
\label{sec:results}

\begin{figure}
\centering
\includegraphics[width=\textwidth]{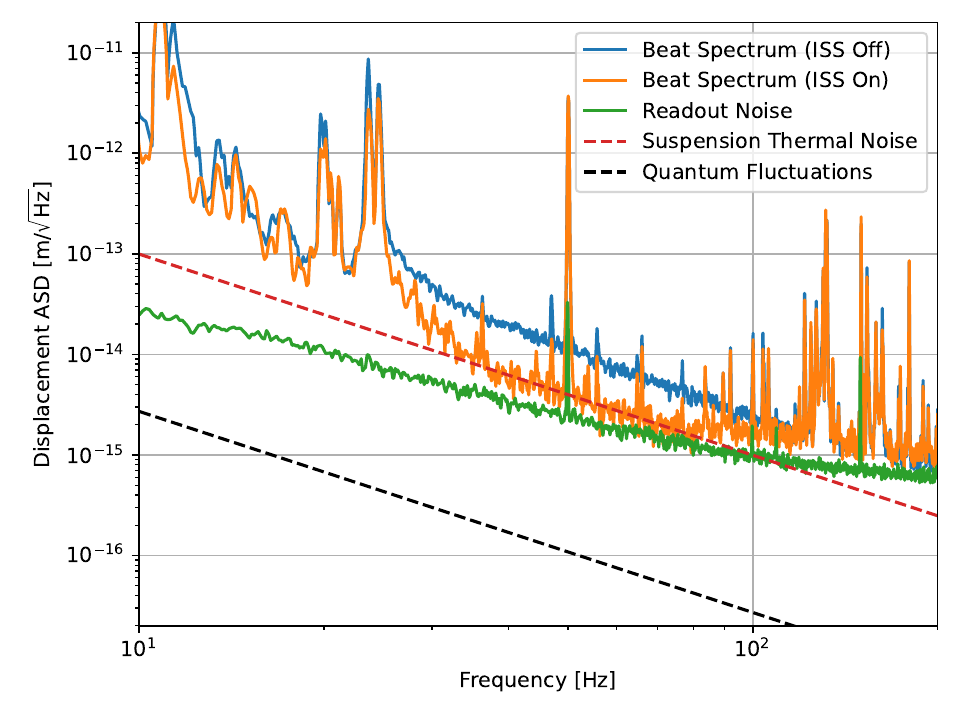}
    \caption{Plot showing the beat spectrum between the reflected fields from the two cavities viewed within the detection band. In the middle band between \SIrange{30}{100}{\Hz}, we observe the effects of laser intensity noise coupling via the classical radiation pressure effect shown in the \textit{ISS off} trace. This noise is stabilised by the ISS down to a combination of the PLL readout noise in the upper region of the detection band, and the suspension thermal noise in the middle region. In the region below \SI{30}{\Hz} we see the seismic noise growing rapidly towards the resonances below \SI{10}{\Hz}, which are shown in Fig.~\ref{subfig:iso-cav}.}
\label{fig:beat_spec}
\end{figure}

We show the amplitude spectral density of the beat between the reflected fields from the two cavities in Fig.~\ref{fig:beat_spec}. The spectrum is shown in two configurations: one with intensity stabilisation servo (ISS) off and one with ISS engaged. In the simplified treatment of the interferometric system, intensity fluctuations couple into the phase-sensitive beat readout as a second-order effect. However, the light cavity mirrors lead to a large mechanical admittance to the classical radiation pressure fluctuations caused by the intensity noise. We show that the suppression of this effect leads to an improvement by up to a factor of 5 in the frequency band from \SIrange{30}{100}{\Hz}.

The seismic and suspension thermal noise is sharply suppressed above the suspension resonances, leaving a broad region from 40 Hz onwards that is dominated by the readout noise. This noise has been shaped by an analogue whitening filter to selectively reduce the readout noise in this critical band without reducing the range at low frequencies, which is needed to handle the high RMS motion arising from the seismic peaks. Currently, the readout noise is limited by the ADC noise. If we successfully improve on the ADC noise, we can achieve a factor of 5 reduction in the readout noise before we become limited by the PLL noise instead.

The key issue with reducing readout noise further is the large RMS in the `quiet' band arising from broad peaks in the \SIrange{200}{400}{\Hz} region. The position of these peaks within the region amplified by the whitening filter means that we saturate the ADC if we attempt to amplify the input and reach the PLL noise. We identify these peaks as acoustic noise coupling to the vibrations of the view-ports of the cryostat's vacuum lid. We mitigate this noise source by minimising acoustic noise in the immediate experimental environment and by wrapping the cryostat in acoustically isolating foam. Whilst we are able to reduce this noise by approximately an order-of-magnitude compared to the original setup, additional isolation will be necessary to reduce the RMS from this region and allow us to reduce the ADC noise further.

At the current level of performance, we are capable of reaching a peak sensitivity of \SI{0.5}{\fmasd} across a broad region in the acoustic frequency band. According to the projected noise budget for ultimate room-temperature operation (Fig.~4.7 in Ref.~\cite{SmetanaThesis}), we will need to achieve a sensitivity of approximately \SI{0.01}{\fmasd} at \SI{100}{\Hz} to reach QRPN-dominated operation and observe the quantum back-action effect in the suspended cavity. At this particular frequency, representing the key frequency of interest for our future SQL measurement that we show in Fig.~\ref{fig:beat_spec}, we reach a sensitivity of \SI{2e-15}{\masd}.

As a point of comparison, the best sensitivities in suspended interferometers have been demonstrated in large-scale facilities. For example, during the O3 observing run, aLIGO demonstrated a sensitivity of \SI{2e-20}{\masd} at \SI{100}{\Hz} in their \SI{4}{\km} interferometer~\cite{LIGOSensO3}. Recent results from the AEI 10-m prototype show a sensitivity of \SI{1e-17}{\masd}~\cite{AEIThesis} at \SI{100}{\Hz}. In a more comparable experiment consisting of light optics in a table-top set-up, a sensitivity at \SI{100}{\Hz} of \SI{2e-12}{\masd} has been demonstrated in the suspended Michelson interferometer of Ref.~\cite{SusMichelson} and \SI{4e-16}{\masd} was demonstrated in the Fabry-Perot Michelson interferometer in Ref.~\cite{Wipf}. In many of these cases, the sensitivity continues to improve towards higher frequencies (e.g. Ref.~\cite{Wipf} reaches \SI{2e-17}{\masd} at \SI{1}{\kHz}), which is due to their dominant noise source, thermal noise, improving towards higher frequencies, whilst our device is currently limited by the flat noise profile of the readout electronics above \SI{100}{\Hz}.

\section{Conclusion}
\label{sec:conc}

The quantum nature of light, coupled with the excellent sensitivity of modern interferometric devices, opens the door to the development of novel sensors for probing quantum mechanics on the macroscopic scale. In this paper, we present our development of a table-top quantum sensor with femtometre sensitivity within the acoustic frequency band.

Our sensor utilises the expertise developed within the gravitational-wave community, particularly in the control of high-finesse, suspended interferometers, and the mitigation of seismic and thermal noise. We construct a system consisting of a cryostat-mounted four-stage suspension supporting two cavities with finesse in excess of \num[print-unity-mantissa=false]{e5}. We use active isolation of the entire cryostat system to achieve an order-of-magnitude of reduction of the seismic noise prior to the passive isolation of the internal suspension. We achieve the femtometre sensitivity using a custom-made phase lock loop.

The sensor in its current iteration serves as the first sensitivity milestone towards reaching the standard quantum limit. The ultimate low-noise mode will be achieved during cryogenic operation, which will require significant changes to the materials to ones more appropriate for good low-temperature performance. The immediate next step is to reach the sensitivity of \SI{1e-17}{\masd} at 100 Hz by achieving better readout noise, which will, in part, involve the mitigation of ambient acoustic noise.

On a longer time-scale, we will prepare for cryogenic operation by switching to more appropriate materials. For the suspension, this will involve making use of materials with much larger thermal conductivity, such as high-purity aluminium, tungsten or copper. A major change to the optical system will be the switch to silicon resonators with novel but less well-developed optical coatings. With these changes, we predict that it will be possible to reach the standard quantum limit within the same acoustic frequency band presented in this paper.

\vspace{6pt} 

\authorcontributions{Conceptualization, D.M.; methodology, J.S., T.Y., V.B., D.M.; software, J.S., T.Y., D.M.; validation, J.S., T.Y., V.B., D.M.; formal analysis, J.S., T.Y., D.M.; investigation, J.S., T.Y, D.M.; resources, V.B., D.M.; data curation, J.S., T.Y, D.M.; writing---original draft preparation, J.S.; writing---review and editing, J.S., T.Y., V.B., D.M.; visualization, J.S., T.Y.; supervision, D.M.; project administration, V.B, D.M.; funding acquisition, D.M. All authors have read and agreed to the published version of the manuscript.}

\funding{This research was funded by STFC grant numbers ST/T006331/1, ST/T006609/1 and ST/W006375/1.}

\acknowledgments{We acknowledge help of Dr Valera Frolov from LIGO Livingston Observatory during his visit to Birmingham in September 2023 and members of the UK Quantum Interferometry collaboration for useful discussions, the support of the Institute for Gravitational Wave Astronomy at the University of Birmingham and STFC Quantum Technology for Fundamental Physics scheme (Grant No. ST/T006331/1, ST/T006609/1 and ST/W006375/1). D.M. is supported by the 2021 Philip Leverhulme Prize. }

\conflictsofinterest{The authors declare no conflicts of interest. The funders had no role in the design of the study; in the collection, analyses, or interpretation of data; in the writing of the manuscript; or in the decision to publish the results.} 



\abbreviations{Abbreviations}{
The following abbreviations are used in this manuscript:\\

\noindent 
\begin{tabular}{@{}ll}
ADC & Analogue-to-digital converter\\
AdV & Advanced Virgo\\
aLIGO & Advanced LIGO\\
BOSEM & Birmingham optical sensor and electromagnetic motor\\
CDS & Control and data system\\
DoF & Degree of freedom\\
FSR & Free spectral range\\
GW & Gravitational wave\\
ISI & Internal seismic isolation\\
ISS & Intensity stabilisation servo\\
PLL & Phase lock loop\\
QRPN & Quantum radiation pressure noise\\
QSN & Quantum shot noise\\
RMS & Root mean square\\
SQL & Standard quantum limit\\
VCO & Voltage-controlled oscillator\\
VOA & Variable optical attenuator\\
\end{tabular}
}

\begin{adjustwidth}{-\extralength}{0cm}

\reftitle{References}
\bibliography{main}

\begin{thebibliography}{999}

\bibitem[Metcalf and van~der Straten(2003)]{AtomTraps}
Metcalf, H.J.; van~der Straten, P.
\newblock Laser cooling and trapping of atoms.
\newblock {\em J. Opt. Soc. Am. B} {\bf 2003}, {\em 20},~887--908.
\newblock {\url{https://doi.org/10.1364/JOSAB.20.000887}}.

\bibitem[Adams et~al.(1994)Adams, Carnal, and Mlynek]{AtomInterf}
Adams, C.; Carnal, O.; Mlynek, J.
\newblock Atom Interferometry; Academic Press,  1994; Vol.~34, {\em Advances In
  Atomic, Molecular, and Optical Physics}, pp. 1--33.
\newblock
  {\url{https://doi.org/https://doi.org/10.1016/S1049-250X(08)60073-7}}.

\bibitem[Badurina et~al.(2020)Badurina, Bentine, Blas, Bongs, Bortoletto,
  Bowcock, Bridges, Bowden, Buchmueller, Burrage, Coleman, Elertas, Ellis,
  Foot, Gibson, Haehnelt, Harte, Hedges, Hobson, Holynski, Jones, Langlois,
  Lellouch, Lewicki, Maiolino, Majewski, Malik, March-Russell, McCabe, Newbold,
  Sauer, Schneider, Shipsey, Singh, Uchida, Valenzuela, van~der Grinten,
  Vaskonen, Vossebeld, Weatherill, and Wilmut]{AION}
Badurina, L.; Bentine, E.; Blas, D.; Bongs, K.; Bortoletto, D.; Bowcock, T.;
  Bridges, K.; Bowden, W.; Buchmueller, O.; Burrage, C.;  et~al.
\newblock AION: an atom interferometer observatory and network.
\newblock {\em Journal of Cosmology and Astroparticle Physics} {\bf 2020}, {\em
  2020},~011.
\newblock {\url{https://doi.org/10.1088/1475-7516/2020/05/011}}.

\bibitem[Aasi et~al.(2015)Aasi, Abbott, Abbott, et~al.]{AdvLIGO}
Aasi, J.; Abbott, B.P.; Abbott, R.;  et~al.
\newblock Advanced {LIGO}.
\newblock {\em Classical and Quantum Gravity} {\bf 2015}, {\em 32},~074001.
\newblock {\url{https://doi.org/10.1088/0264-9381/32/7/074001}}.

\bibitem[Acernese et~al.(2015)Acernese, Agathos, Agatsuma, et~al.]{AdvVirgo}
Acernese, F.; Agathos, M.; Agatsuma, K.;  et~al.
\newblock Advanced Virgo: a second-generation interferometric gravitational
  wave detector.
\newblock {\em Classical and Quantum Gravity} {\bf 2015}, {\em 32},~024001.
\newblock {\url{https://doi.org/10.1088/0264-9381/32/2/024001}}.

\bibitem[Buikema et~al.(2020)Buikema, Cahillane, Mansell, et~al.]{LIGOSensO3}
Buikema, A.; Cahillane, C.; Mansell, G.L.;  et~al.
\newblock Sensitivity and performance of the Advanced LIGO detectors in the
  third observing run.
\newblock {\em Phys. Rev. D} {\bf 2020}, {\em 102},~062003.
\newblock {\url{https://doi.org/10.1103/PhysRevD.102.062003}}.

\bibitem[Abbott et~al.(2016)Abbott, Abbott, Abbott, et~al.]{BBHDetection}
Abbott, B.P.; Abbott, R.; Abbott, T.D.;  et~al.
\newblock Observation of Gravitational Waves from a Binary Black Hole Merger.
\newblock {\em Phys. Rev. Lett.} {\bf 2016}, {\em 116},~061102.
\newblock {\url{https://doi.org/10.1103/PhysRevLett.116.061102}}.

\bibitem[Abbott et~al.(2019)Abbott, Abbott, Abbott, et~al.]{Catalogue1}
Abbott, B.P.; Abbott, R.; Abbott, T.D.;  et~al.
\newblock GWTC-1: A Gravitational-Wave Transient Catalog of Compact Binary
  Mergers Observed by LIGO and Virgo during the First and Second Observing
  Runs.
\newblock {\em Phys. Rev. X} {\bf 2019}, {\em 9},~031040.
\newblock {\url{https://doi.org/10.1103/PhysRevX.9.031040}}.

\bibitem[Abbott et~al.(2021)Abbott, Abbott, Abraham, et~al.]{Catalogue2}
Abbott, R.; Abbott, T.D.; Abraham, S.;  et~al.
\newblock GWTC-2: Compact Binary Coalescences Observed by LIGO and Virgo during
  the First Half of the Third Observing Run.
\newblock {\em Phys. Rev. X} {\bf 2021}, {\em 11},~021053.
\newblock {\url{https://doi.org/10.1103/PhysRevX.11.021053}}.

\bibitem[Collaboration et~al.(2021)Collaboration, the Virgo~Collaboration, the
  KAGRA~Collaboration, Abbott, Abbott, Acernese, Ackley, Adams, Adhikari,
  Adhikari, Adya, Affeldt, Agarwal, Agathos, Agatsuma, Aggarwal, Aguiar,
  Aiello, Ain, Ajith, Akcay, Akutsu, Albanesi, Allocca, Altin, Amato, Anand,
  Anand, Ananyeva, Anderson, Anderson, Ando, Andrade, Andres, Andrić,
  Angelova, Ansoldi, Antelis, Antier, Appert, Arai, Arai, Arai, Araki, Araya,
  Araya, Areeda, Arène, Aritomi, Arnaud, Arogeti, Aronson, Arun, Asada, Asali,
  Ashton, Aso, Assiduo, Aston, Astone, Aubin, Austin, Babak, Badaracco, Bader,
  Badger, Bae, Bae, Baer, Bagnasco, Bai, Baiotti, Baird, Bajpai, Ball,
  Ballardin, Ballmer, Balsamo, Baltus, Banagiri, Bankar, Barayoga, Barbieri,
  Barish, Barker, Barneo, Barone, Barr, Barsotti, Barsuglia, Barta, Bartlett,
  Barton, Bartos, Bassiri, Basti, Bawaj, Bayley, Baylor, Bazzan, Bécsy,
  et~al.]{Catalogue3}
Collaboration, T.L.S.; the Virgo~Collaboration.; the KAGRA~Collaboration.;
  Abbott, R.; Abbott, T.D.; Acernese, F.; Ackley, K.; Adams, C.; Adhikari, N.;
  Adhikari, R.X.;  et~al.
\newblock GWTC-3: Compact Binary Coalescences Observed by LIGO and Virgo During
  the Second Part of the Third Observing Run,  2021,
  \href{http://arxiv.org/abs/2111.03606}{{\normalfont
  [arXiv:gr-qc/2111.03606]}}.

\bibitem[Mehta et~al.(2023)Mehta, Buonanno, Cotesta, Ghosh, Sennett, and
  Steinhoff]{GRTest}
Mehta, A.K.; Buonanno, A.; Cotesta, R.; Ghosh, A.; Sennett, N.; Steinhoff, J.
\newblock Tests of general relativity with gravitational-wave observations
  using a flexible theory-independent method.
\newblock {\em Phys. Rev. D} {\bf 2023}, {\em 107},~044020.
\newblock {\url{https://doi.org/10.1103/PhysRevD.107.044020}}.

\bibitem[Abbott et~al.(2021)Abbott, Abbott, Abraham, Acernese, Ackley, Adams,
  Adams, Adhikari, Adya, Affeldt, Agathos, Agatsuma, Aggarwal, Aguiar, Aiello,
  Ain, Ajith, Allen, Allocca, Altin, Amato, Anand, Ananyeva, Anderson,
  Anderson, Angelova, Ansoldi, Antelis, Antier, Appert, Arai, Araya, Areeda,
  Arène, Arnaud, Aronson, Arun, Asali, Ascenzi, Ashton, Aston, Astone, Aubin,
  Aufmuth, AultONeal, Austin, Avendano, Babak, Badaracco, Bader, Bae, Baer,
  Bagnasco, Baird, Ball, et~al.]{PopStats}
Abbott, R.; Abbott, T.D.; Abraham, S.; Acernese, F.; Ackley, K.; Adams, A.;
  Adams, C.; Adhikari, R.X.; Adya, V.B.; Affeldt, C.;  et~al.
\newblock Population Properties of Compact Objects from the Second LIGO–Virgo
  Gravitational-Wave Transient Catalog.
\newblock {\em The Astrophysical Journal Letters} {\bf 2021}, {\em 913},~L7.
\newblock {\url{https://doi.org/10.3847/2041-8213/abe949}}.

\bibitem[Abbott et~al.(2018)Abbott, Abbott, Abbott, Acernese, Ackley, Adams,
  Adams, Addesso, Adhikari, Adya, Affeldt, Agarwal, Agathos, Agatsuma,
  Aggarwal, Aguiar, Aiello, Ain, Ajith, Allen, Allen, Allocca, Aloy, Altin,
  Amato, Ananyeva, Anderson, Anderson, Angelova, Antier, Appert, Arai, Araya,
  Areeda, Ar\`ene, Arnaud, Arun, Ascenzi, Ashton, Ast, Aston, Astone, Atallah,
  Aubin, Aufmuth, Aulbert, AultONeal, Austin, Avila-Alvarez, Babak, Bacon,
  Badaracco, Bader, Bae, Baker, Baldaccini, Ballardin, Ballmer, Banagiri,
  Barayoga, Barclay, Barish, Barker, Barkett, Barnum, Barone, Barr, Barsotti,
  et~al.]{NSProperties}
Abbott, B.P.; Abbott, R.; Abbott, T.D.; Acernese, F.; Ackley, K.; Adams, C.;
  Adams, T.; Addesso, P.; Adhikari, R.X.; Adya, V.B.;  et~al.
\newblock GW170817: Measurements of Neutron Star Radii and Equation of State.
\newblock {\em Phys. Rev. Lett.} {\bf 2018}, {\em 121},~161101.
\newblock {\url{https://doi.org/10.1103/PhysRevLett.121.161101}}.

\bibitem[Reitze et~al.(2019)Reitze, Adhikari, Ballmer, et~al.]{CE}
Reitze, D.; Adhikari, R.X.; Ballmer, S.;  et~al.
\newblock Cosmic Explorer: The U.S. Contribution to Gravitational-Wave
  Astronomy beyond LIGO,  2019,
  \href{http://arxiv.org/abs/1907.04833}{{\normalfont
  [arXiv:astro-ph.IM/1907.04833]}}.

\bibitem[{ET Steering Committee Editorial Team}(2020)]{ET2020}
{ET Steering Committee Editorial Team}.
\newblock {\em Einstein Telescope design report update 2020}; Einstein
  Telescope Collaboration,  2020.

\bibitem[Amaro-Seoane et~al.(2017)Amaro-Seoane, Audley, Babak, et~al.]{LISA}
Amaro-Seoane, P.; Audley, H.; Babak, S.;  et~al.
\newblock Laser Interferometer Space Antenna.
\newblock {\em ArXiv} {\bf 2017},
  \href{http://arxiv.org/abs/1702.00786}{{\normalfont
  [arXiv:astro-ph.IM/1702.00786]}}.

\bibitem[Bähre et~al.(2013)Bähre, Döbrich, Dreyling-Eschweiler, Ghazaryan,
  Hodajerdi, Horns, Januschek, Knabbe, Lindner, Notz, Ringwald, von Seggern,
  Stromhagen, Trines, and Willke]{ALPS}
Bähre, R.; Döbrich, B.; Dreyling-Eschweiler, J.; Ghazaryan, S.; Hodajerdi,
  R.; Horns, D.; Januschek, F.; Knabbe, E.A.; Lindner, A.; Notz, D.;  et~al.
\newblock Any light particle search II — Technical Design Report.
\newblock {\em Journal of Instrumentation} {\bf 2013}, {\em 8},~T09001.
\newblock {\url{https://doi.org/10.1088/1748-0221/8/09/T09001}}.

\bibitem[Oshima et~al.(2023)Oshima, Fujimoto, Kume, Morisaki, Nagano, Fujita,
  Obata, Nishizawa, Michimura, and Ando]{DANCE}
Oshima, Y.; Fujimoto, H.; Kume, J.; Morisaki, S.; Nagano, K.; Fujita, T.;
  Obata, I.; Nishizawa, A.; Michimura, Y.; Ando, M.
\newblock First Results of Axion Dark Matter Search with DANCE,  2023,
  \href{http://arxiv.org/abs/2303.03594}{{\normalfont
  [arXiv:hep-ex/2303.03594]}}.

\bibitem[Heinze et~al.(2023)Heinze, Gill, Dmitriev, Smetana, Yan, Boyer,
  Martynov, and Evans]{LIDA}
Heinze, J.; Gill, A.; Dmitriev, A.; Smetana, J.; Yan, T.; Boyer, V.; Martynov,
  D.; Evans, M.
\newblock First results of the Laser-Interferometric Detector for Axions
  (LIDA),  2023,  \href{http://arxiv.org/abs/2307.01365}{{\normalfont
  [arXiv:astro-ph.CO/2307.01365]}}.

\bibitem[Chou et~al.(2016)Chou, Gustafson, Hogan, Kamai, Kwon, Lanza, McCuller,
  Meyer, Richardson, Stoughton, Tomlin, Waldman, and Weiss]{Holometer1}
Chou, A.S.; Gustafson, R.; Hogan, C.; Kamai, B.; Kwon, O.; Lanza, R.; McCuller,
  L.; Meyer, S.S.; Richardson, J.; Stoughton, C.;  et~al.
\newblock First Measurements of High Frequency Cross-Spectra from a Pair of
  Large Michelson Interferometers.
\newblock {\em Phys. Rev. Lett.} {\bf 2016}, {\em 117},~111102.
\newblock {\url{https://doi.org/10.1103/PhysRevLett.117.111102}}.

\bibitem[Richardson et~al.(2021)Richardson, Kwon, Gustafson, Hogan, Kamai,
  McCuller, Meyer, Stoughton, Tomlin, and Weiss]{Holometer2}
Richardson, J.W.; Kwon, O.; Gustafson, H.R.; Hogan, C.; Kamai, B.L.; McCuller,
  L.P.; Meyer, S.S.; Stoughton, C.; Tomlin, R.E.; Weiss, R.
\newblock Interferometric Constraints on Spacelike Coherent Rotational
  Fluctuations.
\newblock {\em Phys. Rev. Lett.} {\bf 2021}, {\em 126},~241301.
\newblock {\url{https://doi.org/10.1103/PhysRevLett.126.241301}}.

\bibitem[Pan et~al.(2012)Pan, Chen, Lu, Weinfurter, Zeilinger, and
  \ifmmode~\dot{Z}\else \.{Z}\fi{}ukowski]{Entanglement}
Pan, J.W.; Chen, Z.B.; Lu, C.Y.; Weinfurter, H.; Zeilinger, A.;
  \ifmmode~\dot{Z}\else \.{Z}\fi{}ukowski, M.
\newblock Multiphoton entanglement and interferometry.
\newblock {\em Rev. Mod. Phys.} {\bf 2012}, {\em 84},~777--838.
\newblock {\url{https://doi.org/10.1103/RevModPhys.84.777}}.

\bibitem[Caves and Schumaker(1985)]{TwoPhoton}
Caves, C.M.; Schumaker, B.L.
\newblock New formalism for two-photon quantum optics. I. Quadrature phases and
  squeezed states.
\newblock {\em Phys. Rev. A} {\bf 1985}, {\em 31},~3068--3092.
\newblock {\url{https://doi.org/10.1103/PhysRevA.31.3068}}.

\bibitem[Schumaker and Caves(1985)]{Schumaker1985New}
Schumaker, B.L.; Caves, C.M.
\newblock New formalism for two-photon quantum optics. II. Mathematical
  foundation and compact notation.
\newblock {\em Phys. Rev. A} {\bf 1985}, {\em 31},~3093--3111.
\newblock {\url{https://doi.org/10.1103/PhysRevA.31.3093}}.

\bibitem[Caves(1980)]{RadPresSource}
Caves, C.M.
\newblock Quantum-Mechanical Radiation-Pressure Fluctuations in an
  Interferometer.
\newblock {\em Phys. Rev. Lett.} {\bf 1980}, {\em 45},~75--79.
\newblock {\url{https://doi.org/10.1103/PhysRevLett.45.75}}.

\bibitem[Miao and Chen(2012)]{QuantumBookMiao}
Miao, H.; Chen, Y.
\newblock {Quantum theory of laser interferometer gravitational wave
  detectors}. In {\em Advanced Gravitational Wave Detectors}; Blair, D.G.; Ju,
  L.; Zhao, C.; Howell, E.J., Eds.; Cambridge University Press: Cambridge,
  2012; pp. 277--297.
\newblock {\url{https://doi.org/10.1017/CBO9781139046916.018}}.

\bibitem[Braginsky(1968)]{BraginskyQuantum}
Braginsky, V.B.
\newblock Classical and Quantum Restrictions on the Detection of Weak
  Disturbances of a Macroscopic Oscillator.
\newblock {\em Journal of Experimental and Theoretical Physics} {\bf 1968},
  {\em 26}.

\bibitem[Caves(1981)]{SqueezingIntroduction}
Caves, C.M.
\newblock {Quantum-mechanical noise in an interferometer}.
\newblock {\em Physical Review D} {\bf 1981}, {\em 23},~1693--1708.
\newblock {\url{https://doi.org/10.1103/PhysRevD.23.1693}}.

\bibitem[Schnabel(2017)]{SqueezingExp1}
Schnabel, R.
\newblock Squeezed states of light and their applications in laser
  interferometers.
\newblock {\em Physics Reports} {\bf 2017}, {\em 684},~1--51.
\newblock Squeezed states of light and their applications in laser
  interferometers,
  {\url{https://doi.org/https://doi.org/10.1016/j.physrep.2017.04.001}}.

\bibitem[Südbeck et~al.(2020)Südbeck, Steinlechner, Korobko, and
  Schnabel]{SqueezingExp2}
Südbeck, J.; Steinlechner, S.; Korobko, M.; Schnabel, R.
\newblock Demonstration of interferometer enhancement through
  Einstein-Podolsky-Rosen entanglement.
\newblock {\em Nature Photonics} {\bf 2020}, {\em 14},~240--244.

\bibitem[McCuller et~al.(2020)McCuller, Whittle, Ganapathy, Komori, Tse,
  Fernandez-Galiana, Barsotti, Fritschel, MacInnis, Matichard, Mason,
  Mavalvala, Mittleman, Yu, Zucker, and Evans]{SqueezingFreqDep}
McCuller, L.; Whittle, C.; Ganapathy, D.; Komori, K.; Tse, M.;
  Fernandez-Galiana, A.; Barsotti, L.; Fritschel, P.; MacInnis, M.; Matichard,
  F.;  et~al.
\newblock Frequency-Dependent Squeezing for Advanced LIGO.
\newblock {\em Phys. Rev. Lett.} {\bf 2020}, {\em 124},~171102.
\newblock {\url{https://doi.org/10.1103/PhysRevLett.124.171102}}.

\bibitem[Aasi et~al.(2013)Aasi, Abadie, Abbott, Abbott, Abbott, Abernathy,
  Adams, Adams, Addesso, Adhikari, et~al.]{SqueezingLIGO}
Aasi, J.; Abadie, J.; Abbott, B.P.; Abbott, R.; Abbott, T.D.; Abernathy, M.R.;
  Adams, C.; Adams, T.; Addesso, P.; Adhikari, R.X.;  et~al.
\newblock Enhanced sensitivity of the LIGO gravitational wave detector by using
  squeezed states of light.
\newblock {\em Nature Photonics} {\bf 2013}, {\em 7},~613--619.

\bibitem[Acernese et~al.(2019)Acernese, Agathos, Aiello, Allocca, Amato,
  Ansoldi, Antier, Ar\`ene, Arnaud, Ascenzi, Astone, Aubin, Babak, Bacon,
  Badaracco, Bader, Baird, Baldaccini, Ballardin, Baltus, Barbieri, Barneo,
  Barone, Barsuglia, Barta, Basti, Bawaj, Bazzan, Bejger, Belahcene, Bernuzzi,
  Bersanetti, Bertolini, Bischi, Bitossi, Bizouard, Bobba, Boer, Bogaert,
  Bondu, Bonnand, Boom, Boschi, Bouffanais, Bozzi, Bradaschia, Branchesi,
  Breschi, Briant, Brighenti, Brillet, Brooks, Bruno, Bulik, Bulten, Buskulic,
  Cagnoli, Calloni, Canepa, Carapella, Carbognani, Carullo, Casanueva~Diaz,
  Casentini, Casta\~neda, Caudill, Cavalier, Cavalieri, Cella, Cerd\'a-Dur\'an,
  Cesarini, Chaibi, Chassande-Mottin, Chiadini, Chierici, Chincarini, Chiummo,
  Christensen, Chua, Ciani, Ciecielag, Cie\ifmmode~\acute{s}\else
  \'{s}\fi{}lar, Ciolfi, Cipriano, Cirone, Clesse, Cleva, Coccia, Cohadon,
  Cohen, Colpi, Conti, Cordero-Carri\'on, Corezzi, Corre, Cortese, Coulon,
  Croquette, Cudell, Cuoco, Curylo, D'Angelo, D'Antonio, Dattilo, Davier,
  Degallaix, De~Laurentis, Del\'eglise, Del~Pozzo, De~Pietri, De~Rosa,
  De~Rossi, Dietrich, Di~Fiore, Di~Giorgio, Di~Giovanni, Di~Giovanni,
  Di~Girolamo, Di~Lieto, Di~Pace, Di~Palma, Di~Renzo, Drago, Ducoin, Durante,
  D'Urso, Eisenmann, Errico, Estevez, Fafone, Farinon, Feng, Fenyvesi,
  Ferrante, Fidecaro, Fiori, Fiorucci, Fittipaldi, Fiumara, Flaminio, Font,
  Fournier, Frasca, Frasconi, Frey, Fronz\`e, Garufi, Gemme, Genin, Gennai,
  Ghosh, Giacomazzo, Gosselin, Gouaty, Grado, Granata, Greco, Grignani,
  Grimaldi, Grimm, Gruning, Guidi, Guix\'e, Guo, Gupta, Halim, Harder, Harms,
  Heidmann, Heitmann, Hello, Hemming, Hennes, Hinderer, Hofman, Huet, Hui,
  Idzkowski, Iess, Intini, Isac, Jacqmin, Jaranowski, Jonker, Katsanevas,
  K\'ef\'elian, Khan, Khetan, Koekoek, Koley, Kr\'olak, Kutynia, Laghi,
  Lamberts, La~Rosa, Lartaux-Vollard, Lazzaro, Leaci, Leroy, Letendre, Linde,
  Llorens-Monteagudo, Longo, Lorenzini, Loriette, Losurdo, Lumaca, Macquet,
  Majorana, Maksimovic, Man, Mangano, Mantovani, Mapelli, Marchesoni, Marion,
  Marquina, Marsat, Martelli, Martinez, Masserot, Mastrogiovanni, Mejuto~Villa,
  Mereni, Merzougui, Metzdorff, Miani, Michel, Milano, Miller, Milotti,
  Minazzoli, Minenkov, Montani, Morawski, Mours, Muciaccia, Nagar, Nardecchia,
  Naticchioni, Neilson, Nelemans, Nguyen, Nichols, Nissanke, Nocera, Oganesyan,
  Olivetto, Pagano, Pagliaroli, Palomba, Pang, Pannarale, Paoletti, Paoli,
  Pascucci, Pasqualetti, Passaquieti, Passuello, Patricelli, Perego, Pegoraro,
  P\'erigois, Perreca, Perri\`es, Phukon, Piccinni, Pichot, Piendibene,
  Piergiovanni, Pierro, Pillant, Pinard, Pinto, Piotrzkowski, Plastino,
  Poggiani, Popolizio, Porter, Prevedelli, Principe, Prodi, Punturo, Puppo,
  Raaijmakers, Radulesco, Rapagnani, Razzano, Regimbau, Rei, Rettegno, Ricci,
  Riemenschneider, Robinet, Rocchi, Rolland, Romanelli, Romano,
  Rosi\ifmmode~\acute{n}\else \'{n}\fi{}ska, Ruggi, Salafia, Salconi, Samajdar,
  Sanchis-Gual, Santos, Sassolas, Sauter, Sayah, Sentenac, Sequino, Sharma,
  Sieniawska, Singh, Singhal, Sipala, Sordini, Sorrentino, Spera, Stachie,
  Steer, Stratta, Sur, Swinkels, Tacca, Tanasijczuk, Tapia San~Martin, Tiwari,
  Tonelli, Torres-Forn\'e, Tosta~e Melo, Travasso, Tringali, Trovato, Tsang,
  Turconi, Valentini, van Bakel, van Beuzekom, van~den Brand, Van Den~Broeck,
  van~der Schaaf, Vardaro, Vas\'uth, Vedovato, Verkindt, Vetrano, Vicer\'e,
  Vinet, Vocca, Walet, Was, Zadro\ifmmode~\dot{z}\else \.{z}\fi{}ny, Zelenova,
  Zendri, Vahlbruch, Mehmet, L\"uck, and Danzmann]{SqueezingVirgo}
Acernese, F.; Agathos, M.; Aiello, L.; Allocca, A.; Amato, A.; Ansoldi, S.;
  Antier, S.; Ar\`ene, M.; Arnaud, N.; Ascenzi, S.;  et~al.
\newblock Increasing the Astrophysical Reach of the Advanced Virgo Detector via
  the Application of Squeezed Vacuum States of Light.
\newblock {\em Phys. Rev. Lett.} {\bf 2019}, {\em 123},~231108.
\newblock {\url{https://doi.org/10.1103/PhysRevLett.123.231108}}.

\bibitem[Bentley et~al.(2019)Bentley, Jones, Martynov, Freise, and
  Miao]{QuantumAmpOriginal}
Bentley, J.; Jones, P.; Martynov, D.; Freise, A.; Miao, H.
\newblock Converting the signal-recycling cavity into an unstable
  optomechanical filter to enhance the detection bandwidth of
  gravitational-wave detectors.
\newblock {\em Phys. Rev. D} {\bf 2019}, {\em 99},~102001.
\newblock {\url{https://doi.org/10.1103/PhysRevD.99.102001}}.

\bibitem[Li et~al.(2021)Li, Smetana, Ubhi, Bentley, Chen, Ma, Miao, and
  Martynov]{QuantumAmplifierTD}
Li, X.; Smetana, J.; Ubhi, A.S.; Bentley, J.; Chen, Y.; Ma, Y.; Miao, H.;
  Martynov, D.
\newblock Enhancing interferometer sensitivity without sacrificing bandwidth
  and stability: Beyond single-mode and resolved-sideband approximation.
\newblock {\em Phys. Rev. D} {\bf 2021}, {\em 103},~122001.
\newblock {\url{https://doi.org/10.1103/PhysRevD.103.122001}}.

\bibitem[Dmitriev et~al.(2022)Dmitriev, Miao, and Martynov]{QuantumAmpIdeal}
Dmitriev, A.; Miao, H.; Martynov, D.
\newblock Enhancing the sensitivity of interferometers with stable
  phase-insensitive quantum filters.
\newblock {\em Phys. Rev. D} {\bf 2022}, {\em 106},~022007.
\newblock {\url{https://doi.org/10.1103/PhysRevD.106.022007}}.

\bibitem[Bersanetti et~al.(2021)Bersanetti, Patricelli, Piccinni, Piergiovanni,
  Salemi, and Sequino]{VirgoSens}
Bersanetti, D.; Patricelli, B.; Piccinni, O.J.; Piergiovanni, F.; Salemi, F.;
  Sequino, V.
\newblock Advanced Virgo: Status of the Detector, Latest Results and Future
  Prospects.
\newblock {\em Universe} {\bf 2021}, {\em 7}.
\newblock {\url{https://doi.org/10.3390/universe7090322}}.

\bibitem[M\"uller-Ebhardt et~al.(2008)M\"uller-Ebhardt, Rehbein, Schnabel,
  Danzmann, and Chen]{MacroscopicEntanglement}
M\"uller-Ebhardt, H.; Rehbein, H.; Schnabel, R.; Danzmann, K.; Chen, Y.
\newblock Entanglement of Macroscopic Test Masses and the Standard Quantum
  Limit in Laser Interferometry.
\newblock {\em Phys. Rev. Lett.} {\bf 2008}, {\em 100},~013601.
\newblock {\url{https://doi.org/10.1103/PhysRevLett.100.013601}}.

\bibitem[Yang et~al.(2013)Yang, Miao, Lee, Helou, and Chen]{SemiGrav}
Yang, H.; Miao, H.; Lee, D.S.; Helou, B.; Chen, Y.
\newblock Macroscopic Quantum Mechanics in a Classical Spacetime.
\newblock {\em Phys. Rev. Lett.} {\bf 2013}, {\em 110},~170401.
\newblock {\url{https://doi.org/10.1103/PhysRevLett.110.170401}}.

\bibitem[Miao et~al.(2020)Miao, Martynov, Yang, and Datta]{QuantumGrav}
Miao, H.; Martynov, D.; Yang, H.; Datta, A.
\newblock Quantum correlations of light mediated by gravity.
\newblock {\em Phys. Rev. A} {\bf 2020}, {\em 101},~063804.
\newblock {\url{https://doi.org/10.1103/PhysRevA.101.063804}}.

\bibitem[Rao and Culshaw(1992)]{Microresonators}
Rao, Y.; Culshaw, B.
\newblock Comparison between optically excited vibrations of silicon cantilever
  and bridge microresonators.
\newblock {\em Sensors and Actuators A: Physical} {\bf 1992}, {\em
  30},~203--208.
\newblock {\url{https://doi.org/https://doi.org/10.1016/0924-4247(92)80121-I}}.

\bibitem[Sidles and Sigg(2006)]{SiggSidles}
Sidles, J.A.; Sigg, D.
\newblock Optical torques in suspended Fabry–Perot interferometers.
\newblock {\em Physics Letters A} {\bf 2006}, {\em 354},~167--172.
\newblock
  {\url{https://doi.org/https://doi.org/10.1016/j.physleta.2006.01.051}}.

\bibitem[Wipf(2013)]{Wipf}
Wipf, C.
\newblock Toward Quantum Opto-Mechanics in a Gram-Scale Suspended Mirror
  Interferometer.
\newblock PhD thesis, Massachusetts Institute of Technology,  2013.

\bibitem[Westphal et~al.(2012)Westphal, Bergmann, Bertolini, Born, Chen,
  Cumming, Cunningham, Dahl, Gr{\"{a}}f, Hammond, Heinzel, Hild, Huttner,
  Jones, Kawazoe, K{\"{o}}hlenbeck, K{\"{u}}hn, L{\"{u}}ck, Mossavi,
  P{\"{o}}ld, Somiya, van Veggel, Wanner, Willke, Strain, Go{\ss}ler, and
  Danzmann]{AEI10m}
Westphal, T.; Bergmann, G.; Bertolini, A.; Born, M.; Chen, Y.; Cumming, A.V.;
  Cunningham, L.; Dahl, K.; Gr{\"{a}}f, C.; Hammond, G.;  et~al.
\newblock {Design of the 10 m AEI prototype facility for interferometry
  studies}.
\newblock {\em Applied Physics B} {\bf 2012}, {\em 106},~551--557.
\newblock {\url{https://doi.org/10.1007/s00340-012-4878-z}}.

\bibitem[Smetana and Martynov(2020)]{ConferenceSQL1}
Smetana, J.; Martynov, D.
\newblock Towards the Standard Quantum Limit in a Table-Top Interferometer.
\newblock In Proceedings of the Proceedings of the GRavitational-waves
  Science\&technology Symposium (GRASS) 2019. Zenodo,  2020.
\newblock {\url{https://doi.org/10.5281/zenodo.3561043}}.

\bibitem[Smetana et~al.(2023)Smetana, Yan, Boyer, and Martynov]{ConferenceSQL2}
Smetana, J.; Yan, T.; Boyer, V.; Martynov, D.
\newblock {Reaching a macroscopic quantum state in a suspended interferometer}.
\newblock In Proceedings of the Quantum Technology: Driving Commercialisation
  of an Enabling Science III; Padgett, M.J.; Bongs, K.; Fedrizzi, A.; Politi,
  A., Eds. International Society for Optics and Photonics, SPIE,  2023, Vol.
  12335, p. 1233504.
\newblock {\url{https://doi.org/10.1117/12.2645113}}.

\bibitem[Smetana(2023)]{SmetanaThesis}
Smetana, J.
\newblock Applications of Precision Interferometry in Quantum Measurement and
  Gravitational Wave Detection.
\newblock PhD thesis, Institute for Gravitational Wave Astronomy, University of
  Birmingham,  2023.

\bibitem[Callen and Welton(1951)]{FlucDiss1}
Callen, H.B.; Welton, T.A.
\newblock {Irreversibility and Generalized Noise}.
\newblock {\em Physical Review} {\bf 1951}, {\em 83},~34--40.
\newblock {\url{https://doi.org/10.1103/PhysRev.83.34}}.

\bibitem[Greene and Callen(1951)]{FlucDiss2}
Greene, R.F.; Callen, H.B.
\newblock On the Formalism of Thermodynamic Fluctuation Theory.
\newblock {\em Phys. Rev.} {\bf 1951}, {\em 83},~1231--1235.
\newblock {\url{https://doi.org/10.1103/PhysRev.83.1231}}.

\bibitem[Callen and Greene(1952)]{FlucDiss3}
Callen, H.B.; Greene, R.F.
\newblock {On a Theorem of Irreversible Thermodynamics}.
\newblock {\em Physical Review} {\bf 1952}, {\em 86},~702--710.
\newblock {\url{https://doi.org/10.1103/PhysRev.86.702}}.

\bibitem[Saulson(1990)]{Saulson}
Saulson, P.R.
\newblock {Thermal noise in mechanical experiments}.
\newblock {\em Physical Review D} {\bf 1990}, {\em 42},~2437--2445.
\newblock {\url{https://doi.org/10.1103/PhysRevD.42.2437}}.

\bibitem[Akutsu et~al.(2019)Akutsu, Ando, Arai, et~al.]{Kagra}
Akutsu, T.; Ando, M.; Arai, K.;  et~al.
\newblock KAGRA: 2.5 generation interferometric gravitational wave detector.
\newblock {\em Nature Astronomy} {\bf 2019}, {\em 3},~35--40.
\newblock {\url{https://doi.org/10.1038/s41550-018-0658-y}}.

\bibitem[Adhikari et~al.(2020)Adhikari, Arai, Brooks, Wipf, Aguiar, Altin,
  Barr, Barsotti, Bassiri, Bell, Billingsley, Birney, Blair, Bonilla, Briggs,
  Brown, Byer, Cao, Constancio, Cooper, Corbitt, Coyne, Cumming, Daw, deRosa,
  Eddolls, Eichholz, Evans, Fejer, Ferreira, Freise, Frolov, Gras, Green,
  Grote, Gustafson, Hall, Hammond, Harms, Harry, Haughian, Heinert, Heintze,
  Hellman, Hennig, Hennig, Hild, Hough, Johnson, Kamai, Kapasi, Komori,
  Koptsov, Korobko, Korth, Kuns, Lantz, Leavey, Magana-Sandoval, Mansell,
  Markosyan, Markowitz, Martin, Martin, Martynov, McClelland, McGhee, McRae,
  Mills, Mitrofanov, Molina-Ruiz, Mow-Lowry, Munch, Murray, Ng, Okada, Ottaway,
  Prokhorov, Quetschke, Reid, Reitze, Richardson, Robie, Romero-Shaw, Route,
  Rowan, Schnabel, Schneewind, Seifert, Shaddock, Shapiro, Shoemaker, Silva,
  Slagmolen, Smith, Smith, Steinlechner, Strain, Taira, Tait, Tanner, Tornasi,
  Torrie, Veggel, Vanheijningen, Veitch, Wade, Wallace, Ward, Weiss, Wessels,
  Willke, Yamamoto, Yap, and Zhao]{Voyager}
Adhikari, R.X.; Arai, K.; Brooks, A.F.; Wipf, C.; Aguiar, O.; Altin, P.; Barr,
  B.; Barsotti, L.; Bassiri, R.; Bell, A.;  et~al.
\newblock A cryogenic silicon interferometer for gravitational-wave detection.
\newblock {\em Classical and Quantum Gravity} {\bf 2020}, {\em 37},~165003.
\newblock {\url{https://doi.org/10.1088/1361-6382/ab9143}}.

\bibitem[Hild et~al.(2006)Hild, L\"{u}ck, Winkler, Strain, Grote, Smith, Malec,
  Hewitson, Willke, Hough, and Danzmann]{SilicaAbsorption}
Hild, S.; L\"{u}ck, H.; Winkler, W.; Strain, K.; Grote, H.; Smith, J.; Malec,
  M.; Hewitson, M.; Willke, B.; Hough, J.;  et~al.
\newblock Measurement of a low-absorption sample of OH-reduced fused silica.
\newblock {\em Appl. Opt.} {\bf 2006}, {\em 45},~7269--7272.
\newblock {\url{https://doi.org/10.1364/AO.45.007269}}.

\bibitem[Numata et~al.(2002)Numata, Otsuka, Ando, and Tsubono]{VariousSilica}
Numata, K.; Otsuka, S.; Ando, M.; Tsubono, K.
\newblock Intrinsic losses in various kinds of fused silica.
\newblock {\em Classical and Quantum Gravity} {\bf 2002}, {\em 19},~1697--1702.
\newblock {\url{https://doi.org/10.1088/0264-9381/19/7/363}}.

\bibitem[Schroeter et~al.(2007)Schroeter, Nawrodt, Schnabel, Reid, Martin,
  Rowan, Schwarz, Koettig, Neubert, Thürk, Vodel, Tünnermann, Danzmann, and
  Seidel]{SilicaPeak}
Schroeter, A.; Nawrodt, R.; Schnabel, R.; Reid, S.; Martin, I.; Rowan, S.;
  Schwarz, C.; Koettig, T.; Neubert, R.; Thürk, M.;  et~al.
\newblock On the mechanical quality factors of cryogenic test masses from fused
  silica and crystalline quartz,  2007.
\newblock {\url{https://doi.org/10.48550/ARXIV.0709.4359}}.

\bibitem[Numata and Yamamoto(2012)]{Cryogenics}
Numata, K.; Yamamoto, K.
\newblock {Cryogenics}. In {\em Optical coatings and thermal noise in precision
  measurement}; Harry, G.; Bodiya, T.P.; DeSalvo, R., Eds.; Cambridge
  University Press: Cambridge,  2012; chapter~8, pp. 108--128.

\bibitem[Lyon et~al.(1977)Lyon, Salinger, Swenson, and White]{SiliconExpansion}
Lyon, K.G.; Salinger, G.L.; Swenson, C.A.; White, G.K.
\newblock Linear thermal expansion measurements on silicon from 6 to 340 K.
\newblock {\em Journal of Applied Physics} {\bf 1977}, {\em 48},~865--868,
  \href{http://arxiv.org/abs/https://doi.org/10.1063/1.323747}{{\normalfont
  [https://doi.org/10.1063/1.323747]}}.
\newblock {\url{https://doi.org/10.1063/1.323747}}.

\bibitem[Sites et~al.(1983)Sites, Gilstrap, and Rujkorakarn]{IBS}
Sites, J.R.; Gilstrap, P.; Rujkorakarn, R.
\newblock {Ion Beam Sputter Deposition Of Optical Coatings}.
\newblock {\em Optical Engineering} {\bf 1983}, {\em 22},~224447.
\newblock {\url{https://doi.org/10.1117/12.7973141}}.

\bibitem[Harry et~al.(2006)Harry, Abernathy, Becerra-Toledo, Armandula, Black,
  Dooley, Eichenfield, Nwabugwu, Villar, Crooks, Cagnoli, Hough, How, MacLaren,
  Murray, Reid, Rowan, Sneddon, Fejer, Route, Penn, Ganau, Mackowski, Michel,
  Pinard, and Remillieux]{Absorption}
Harry, G.M.; Abernathy, M.R.; Becerra-Toledo, A.E.; Armandula, H.; Black, E.;
  Dooley, K.; Eichenfield, M.; Nwabugwu, C.; Villar, A.; Crooks, D.R.M.;
  et~al.
\newblock Titania-doped tantala/silica coatings for gravitational-wave
  detection.
\newblock {\em Classical and Quantum Gravity} {\bf 2006}, {\em 24},~405.
\newblock {\url{https://doi.org/10.1088/0264-9381/24/2/008}}.

\bibitem[Harry et~al.(2002)Harry, Gretarsson, Saulson, Kittelberger, Penn,
  Startin, Rowan, Fejer, Crooks, Cagnoli, Hough, and Nakagawa]{CoatingThermal}
Harry, G.M.; Gretarsson, A.M.; Saulson, P.R.; Kittelberger, S.E.; Penn, S.D.;
  Startin, W.J.; Rowan, S.; Fejer, M.M.; Crooks, D.R.M.; Cagnoli, G.;  et~al.
\newblock {Thermal noise in interferometric gravitational wave detectors due to
  dielectric optical coatings}.
\newblock {\em Classical and Quantum Gravity} {\bf 2002}, {\em 19},~897--917.
\newblock {\url{https://doi.org/10.1088/0264-9381/19/5/305}}.

\bibitem[Evans et~al.(2013)Evans, Barsotti, Kwee, Harms, and
  Miao]{RealisticCavities}
Evans, M.; Barsotti, L.; Kwee, P.; Harms, J.; Miao, H.
\newblock Realistic filter cavities for advanced gravitational wave detectors.
\newblock {\em Phys. Rev. D} {\bf 2013}, {\em 88},~022002.
\newblock {\url{https://doi.org/10.1103/PhysRevD.88.022002}}.

\bibitem[Drever et~al.(1983)Drever, Hall, Kowalski, Hough, Ford, Munley, and
  Ward]{PDHScheme}
Drever, R.W.P.; Hall, J.L.; Kowalski, F.V.; Hough, J.; Ford, G.M.; Munley,
  A.J.; Ward, H.
\newblock Laser phase and frequency stabilization using an optical resonator.
\newblock {\em Applied Physics B} {\bf 1983}, {\em 31},~97--105.

\bibitem[Robertson et~al.(2002)Robertson, Cagnoli, Crooks, Elliffe, Faller,
  Fritschel, Go{\$}szlig{\$}ler, Grant, Heptonstall, Hough, L{\$}uuml{\$}ck,
  Mittleman, Perreur-Lloyd, Plissi, Rowan, Shoemaker, Sneddon, Strain, Torrie,
  Ward, and Willems]{LIGOQuad}
Robertson, N.A.; Cagnoli, G.; Crooks, D.R.M.; Elliffe, E.; Faller, J.E.;
  Fritschel, P.; Go{\$}szlig{\$}ler, S.; Grant, A.; Heptonstall, A.; Hough, J.;
   et~al.
\newblock {Quadruple suspension design for Advanced LIGO}.
\newblock {\em Classical and Quantum Gravity} {\bf 2002}, {\em 19},~311.
\newblock {\url{https://doi.org/10.1088/0264-9381/19/15/311}}.

\bibitem[Braccini et~al.(2005)Braccini, Barsotti, Bradaschia,
  et~al.]{VirgoSuperattenuator}
Braccini, S.; Barsotti, L.; Bradaschia, C.;  et~al.
\newblock {Measurement of the seismic attenuation performance of the VIRGO
  Superattenuator}.
\newblock {\em Astroparticle Physics} {\bf 2005}, {\em 23},~557--565.
\newblock {\url{https://doi.org/10.1016/J.ASTROPARTPHYS.2005.04.002}}.

\bibitem[Plissi et~al.(2000)Plissi, Torrie, Husman, Robertson, Strain, Ward,
  Lück, and Hough]{LocalControl}
Plissi, M.V.; Torrie, C.I.; Husman, M.E.; Robertson, N.A.; Strain, K.A.; Ward,
  H.; Lück, H.; Hough, J.
\newblock GEO 600 triple pendulum suspension system: Seismic isolation and
  control.
\newblock {\em Review of Scientific Instruments} {\bf 2000}, {\em
  71},~2539--2545,
  \href{http://arxiv.org/abs/https://doi.org/10.1063/1.1150645}{{\normalfont
  [https://doi.org/10.1063/1.1150645]}}.
\newblock {\url{https://doi.org/10.1063/1.1150645}}.

\bibitem[Torrie(1999)]{BladeEquationTriangle}
Torrie, C.
\newblock Development of Suspensions for the GEO 600 Gravitational Wave
  Detector.
\newblock PhD thesis, University of Glasgow,  1999.

\bibitem[Lula(1986)]{MaragingYield}
Lula, R.A., Stainless Steel; American Society for Metals,  1986; pp. 37--39.

\bibitem[Plissi et~al.(2004)Plissi, Torrie, Barton, Robertson, Grant, Cantley,
  Strain, Willems, Romie, Skeldon, Perreur-Lloyd, Jones, and
  Hough]{EddyDamping}
Plissi, M.V.; Torrie, C.I.; Barton, M.; Robertson, N.A.; Grant, A.; Cantley,
  C.A.; Strain, K.A.; Willems, P.A.; Romie, J.H.; Skeldon, K.D.;  et~al.
\newblock {An investigation of eddy-current damping of multi-stage pendulum
  suspensions for use in interferometric gravitational wave detectors}.
\newblock {\em Review of Scientific Instruments} {\bf 2004}, {\em
  75},~4516--4522.
\newblock {\url{https://doi.org/10.1063/1.1795192}}.

\bibitem[Carbone et~al.(2012)Carbone, Aston, Cutler, Freise, Greenhalgh,
  Heefner, Hoyland, Lockerbie, Lodhia, Robertson, Speake, Strain, and
  Vecchio]{Bosem}
Carbone, L.; Aston, S.M.; Cutler, R.M.; Freise, A.; Greenhalgh, J.; Heefner,
  J.; Hoyland, D.; Lockerbie, N.A.; Lodhia, D.; Robertson, N.A.;  et~al.
\newblock Sensors and actuators for the Advanced {LIGO} mirror suspensions.
\newblock {\em Classical and Quantum Gravity} {\bf 2012}, {\em 29},~115005.
\newblock {\url{https://doi.org/10.1088/0264-9381/29/11/115005}}.

\bibitem[Dasgupta and Mruthyunjaya(2000)]{StewartPlatform}
Dasgupta, B.; Mruthyunjaya, T.S.
\newblock The Stewart platform manipulator: a review.
\newblock {\em Mechanism and Machine Theory} {\bf 2000}, {\em 35},~15--40.
\newblock
  {\url{https://doi.org/https://doi.org/10.1016/S0094-114X(99)00006-3}}.

\bibitem[Ubhi et~al.(2022)Ubhi, Prokhorov, Cooper, Di~Fronzo, Bryant, Hoyland,
  Mitchell, van Dongen, Mow-Lowry, Cumming, Hammond, and Martynov]{New6D}
Ubhi, A.S.; Prokhorov, L.; Cooper, S.; Di~Fronzo, C.; Bryant, J.; Hoyland, D.;
  Mitchell, A.; van Dongen, J.; Mow-Lowry, C.; Cumming, A.;  et~al.
\newblock Active platform stabilisation with a 6D seismometer,  2022.
\newblock {\url{https://doi.org/10.48550/ARXIV.2207.10417}}.

\bibitem[Matichard et~al.(2015)Matichard, Lantz, Mittleman,
  et~al.]{LIGOSeismicStrategy}
Matichard, F.; Lantz, B.; Mittleman, R.;  et~al.
\newblock {Seismic isolation of Advanced LIGO: Review of strategy,
  instrumentation and performance}.
\newblock {\em Classical and Quantum Gravity} {\bf 2015}, {\em 32},~185003.
\newblock {\url{https://doi.org/10.1088/0264-9381/32/18/185003}}.

\bibitem[Giaime et~al.(1996)Giaime, Saha, Shoemaker, and Sievers]{Viton}
Giaime, J.; Saha, P.; Shoemaker, D.; Sievers, L.
\newblock {A passive vibration isolation stack for LIGO: Design, modeling, and
  testing}.
\newblock {\em Review of Scientific Instruments} {\bf 1996}, {\em
  67},~208--214,
  \href{http://arxiv.org/abs/https://pubs.aip.org/aip/rsi/article-pdf/67/1/208/8810244/208\_1\_online.pdf}{{\normalfont
  [https://pubs.aip.org/aip/rsi/article-pdf/67/1/208/8810244/208\_1\_online.pdf]}}.
\newblock {\url{https://doi.org/10.1063/1.1146573}}.

\bibitem[W\"ohler(2023)]{AEIThesis}
W\"ohler, J.
\newblock Direct measurement of coating thermal noise in the AEI 10m prototype.
\newblock PhD thesis, Gottfried Wilhelm Leibniz Universit\"at,  2023.

\bibitem[Agatsuma et~al.(2014)Agatsuma, Friedrich, Ballmer, DeSalvo, Sakata,
  Nishida, and Kawamura]{SusMichelson}
Agatsuma, K.; Friedrich, D.; Ballmer, S.; DeSalvo, G.; Sakata, S.; Nishida, E.;
  Kawamura, S.
\newblock Precise measurement of laser power using an optomechanical system.
\newblock {\em Opt. Express} {\bf 2014}, {\em 22},~2013--2030.
\newblock {\url{https://doi.org/10.1364/OE.22.002013}}.

\end{thebibliography}
\PublishersNote{}
\end{adjustwidth}
\end{document}